\documentclass[reprint,%
 superscriptaddress,%
 twocolumn,%
 floatfix,
 amssymb,%
 amsmath, 
prb, 
 showkeys,8pt
 ,fleqn
 ] 
 {revtex4-1}
\usepackage{bm}
\usepackage{textgreek}
\usepackage{tabularx} 
\usepackage{amsmath}
\usepackage{mathtools}
\usepackage{graphicx}
\usepackage{latexsym}
\usepackage{lineno}
\usepackage[utf8]{inputenc} 
\usepackage{siunitx}
\usepackage{blindtext}
\usepackage{xcolor}
\usepackage{subcaption}
\usepackage{comment}
\usepackage{float}

\begin{document}
	
\title{Modal analysis for determining the size-and temperature-dependent\\ bending rigidity of graphene}
\author{Banafsheh Sajadi}
\author{Simon van Hemert}
\affiliation{Department of Precision and Microsystem Engineering, Faculty of Mechanical, Maritime and Materials Engineering, 
Delft University of Technology, 2628 CD, Delft, The Netherlands.}
\author{Behrouz Arash}
\affiliation{ Department of Structural Engineering, Faculty of Civil Engineering and Geosciences,
Delft University of Technology, 2628 CN, Delft, The Netherlands.}
\author{Pierpaolo Belardinelli}
\affiliation{Department of Precision and Microsystem Engineering, Faculty of Mechanical, Maritime and Materials Engineering, 
Delft University of Technology, 2628 CD, Delft, The Netherlands.}
\author{Peter G. Steeneken}
\affiliation{Department of Precision and Microsystem Engineering, Faculty of Mechanical, Maritime and Materials Engineering, 
Delft University of Technology, 2628 CD, Delft, The Netherlands.}
\affiliation{Kavli Institute of Nanoscience, Faculty of Applied Sciences,
 Delft University of Technology, 2628 CJ, Delft, The Netherlands.}
\author{Farbod Alijani}\email{corresponding author, f.alijani@tudelft.nl}
\affiliation{Department of Precision and Microsystem Engineering, Faculty of Mechanical, Maritime and Materials Engineering, 
Delft University of Technology, 2628 CD, Delft, The Netherlands.}

\begin{abstract}
The bending rigidity of two-dimensional (2D) materials is a key parameter for understanding the mechanics of 2D NEMS devices.  The apparent bending rigidity of graphene membranes at macroscopic scale differs from theoretical predictions at micro-scale. This difference is believed to originate from thermally induced dynamic ripples in the atomically thin membrane. In this paper,  we perform modal analysis to estimate  the effective macroscopic bending rigidity of  graphene membranes from the frequency spectrum of their Brownian motion. Our method is based on fitting the resonance frequencies obtained from the Brownian motion in molecular dynamics simulations, to those obtained from a continuum mechanics model, with bending rigidity and pretension as the fit parameters. In this way, the effective bending rigidity of the membrane and its temperature and size dependence, are extracted, while including the effects of dynamic ripples and thermal fluctuations. The proposed method provides a framework for estimating the macroscopic mechanical properties at high frequencies in other two-dimensional nano-structures at finite temperatures.
\end{abstract}
\keywords{Characterization; Multi-modal approach; Molecular dynamics; Graphene; Bending rigidity.}
\maketitle
\typeout{Filename: reftest4-1.tex for revtex 4.1i 2009/10/19 (AO)}

\section{Introduction}
\label{intro}
The exceptional  mechanical properties of graphene have made it a promising candidate for the next generation of 2D nano-resonators with potential applications in pressure sensing~\cite{Dolleman2015}, mass sensing~\cite{Atalaya2010, Schedin2007}, and electronics~\cite{Schwierz2010, Changyao2013, Novoselov2012}. Proper understanding of the mechanics of this material is not only of fundamental interest, but also a key step towards the development of new devices. Therefore, the elastic properties of graphene have been  investigated in many theoretical and experimental studies~\cite{Zhao2009,Eckmann2012, Lindahl2012, SajadiJAP2017,Farbod2017}. 

The bending rigidity of graphene, however, is still far from being well-understood and compared to its Young's modulus, it is much less investigated. This is due to the fact that for a single atom thick membrane, this parameter is not determined by layer thickness, but by the bending-induced changes in interactions between electron orbitals. In fact, due to its low bending rigidity as compared to the limit of the continuum plate theories, graphene is commonly assumed to have a membrane-like behavior with a negligible (zero) bending rigidity~\cite{SajadiJAP2017,Farbod2017}. 

Direct measurement of bending rigidity has therefore been challenging for mono-layer graphene, as well as other atomically thin membranes. The mostly cited experimental value of \SI{1.2}{\electronvolt} was derived from the phonon spectrum of graphite~\cite{Nicklow1972}. In another study,~\citet{Lindahl2012} proposed a framework  for extracting the bending rigidity of a graphene membrane from the snap-through behavior of its buckled configuration. Based on the proposed method, the authors reported a bending rigidity of  \SI{7.1}{\electronvolt} with a large uncertainty of (\SI{-3}{\electronvolt} to +\SI{4}{\electronvolt}) for mono-layer graphene. In a more recent study, \citet{Blees2015} measured effective bending rigidity of $10^3$–-$10^4$ \SI{}{\electronvolt}. In this study, the authors suggested significant effects of thermal fluctuations as well as static wrinkles on the obtained large bending rigidity.

On the other hand, many studies have investigated the theoretical limit of the bending rigidity of mono-layer graphene~\cite{Yujie2012, Tersoff1992,lu2009,  Sanchez1999}. The theoretical calculations of the bending rigidity for mono-layer graphene have a large range of \SI{0.69}{\electronvolt}--\SI{0.83}{\electronvolt} by models using the Brenner potentials~\cite{Brenner1990, Brenner2002}, and \SI{1.4}{\electronvolt}--
\SI{1.6}{\electronvolt} by  semi-analytical and  density functional theories~\cite{Koskinen2010,lu2009,Kudin2001,Sanchez1999}. It has been reported that bond-angle effects and the bond associated with the dihedral angles are in fact the two dominant sources of the apparent finite bending rigidity of graphene membranes~\cite{lu2009}.
In addition to these effects, \citet{Katsnelson2011} suggested that the bending rigidity of graphene at finite temperatures is also highly influenced by the thermodynamics. In \cite{Katsnelson2011}, the authors used a self-consistent theory of elastic membranes \cite{Nelson1987} and proposed a thermodynamical approximation for the effective wave vector dependent bending rigidity ($\kappa$) in formation of dynamic ripples:
\begin{equation}
\label{bendingkatsnelson}
\kappa=\kappa_0+k_BTA(q_0/q)^\eta,
\end{equation}
where, $T$ is the temperature, $k_B$ is the Boltzmann constant, $\kappa_0=\SI{1}{\electronvolt}$, $A=5.9T^{(\eta/2-1)}$, 
$\eta=$0.85, $q_0=2\pi\sqrt{E_{2D}/\kappa_0}$~\cite{Nicholl2015,Katsnelson2011}, and $q$ is the wave number associated with dynamic ripples. These ripples are also shown to be large enough to affect the effective macroscopic mechanical properties of atomically thin membranes and ribbons~\cite{Deng2016, Gao2014, bao2009,Nicholl2015,Katsnelson2011,Duanduan2017}.

In this paper, we propose a novel approach based on modal analysis for direct estimation of the macroscopic bending rigidity of  graphene membranes. Our method incorporates the effect of Brownian motion and  the resulting ripples on the bending rigidity. We determine a single bending rigidity and pretension with which our model can accurately reproduce up to 10 vibration modes and natural frequencies obtained from atomistic simulations. Furthermore, we show that our obtained bending rigidity can be best fitted with an effective wave number $q_{\mathrm{eff}}$=$\pi/R$,  from Equation~(\ref{bendingkatsnelson}), where $R$ is the radius of the membrane.

The proposed approach for determining the bending rigidity of graphene is outlined as follows: In Section~\ref{Numeric} we employ Molecular Dynamics (MD) simulations to model the Brownian motion in the graphene membrane at finite temperatures. The natural frequencies of the MD model are obtained by applying Fast Fourier Transform (FFT) to
the time signals extracted from MD. Next, in Section~\ref{identification}, we derive a continuum mechanics (CM)  model for the resonance frequencies of a prestressed circular graphene membrane as a function of its pretension and bending rigidity. Finally, by fitting the resonance frequencies obtained from the Brownian motion, to those obtained from CM, the effective bending rigidity at high frequencies is extracted. Moreover, in Section~\ref{results}, the effects of different temperatures and radii of the membrane on  the bending rigidity are discussed, and the results are compared to Equation (\ref{bendingkatsnelson}). 
\begin{figure}[t]
   \centering
    \begin{subfigure}[b]{1\linewidth}
        \centering
        \includegraphics[scale=0.18]{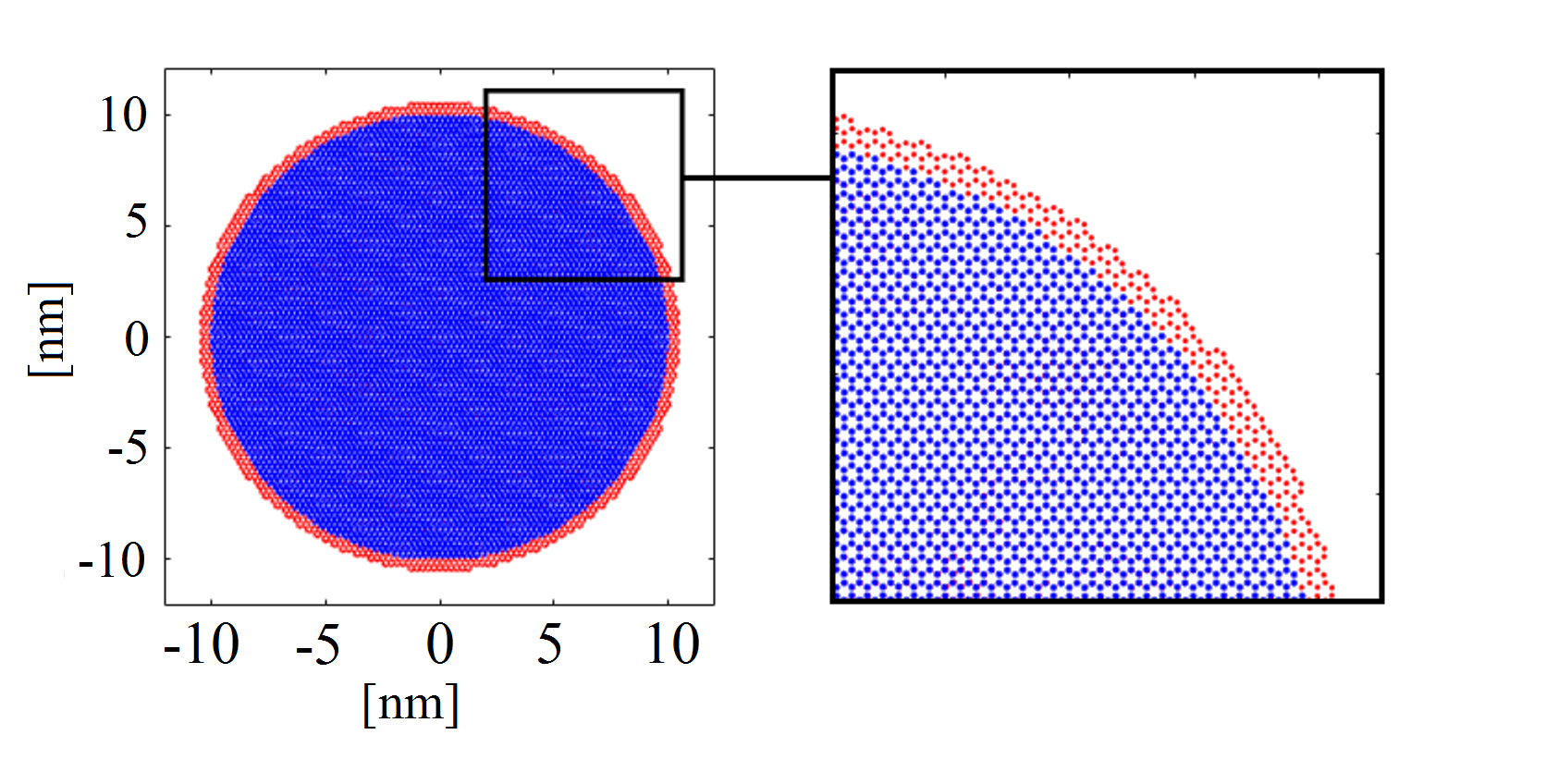}
        \caption{\label{figMDmodel}}
				\vspace{0.3cm}
    \end{subfigure}
    \begin{subfigure}[b]{1\linewidth}
        \includegraphics[scale=0.135]{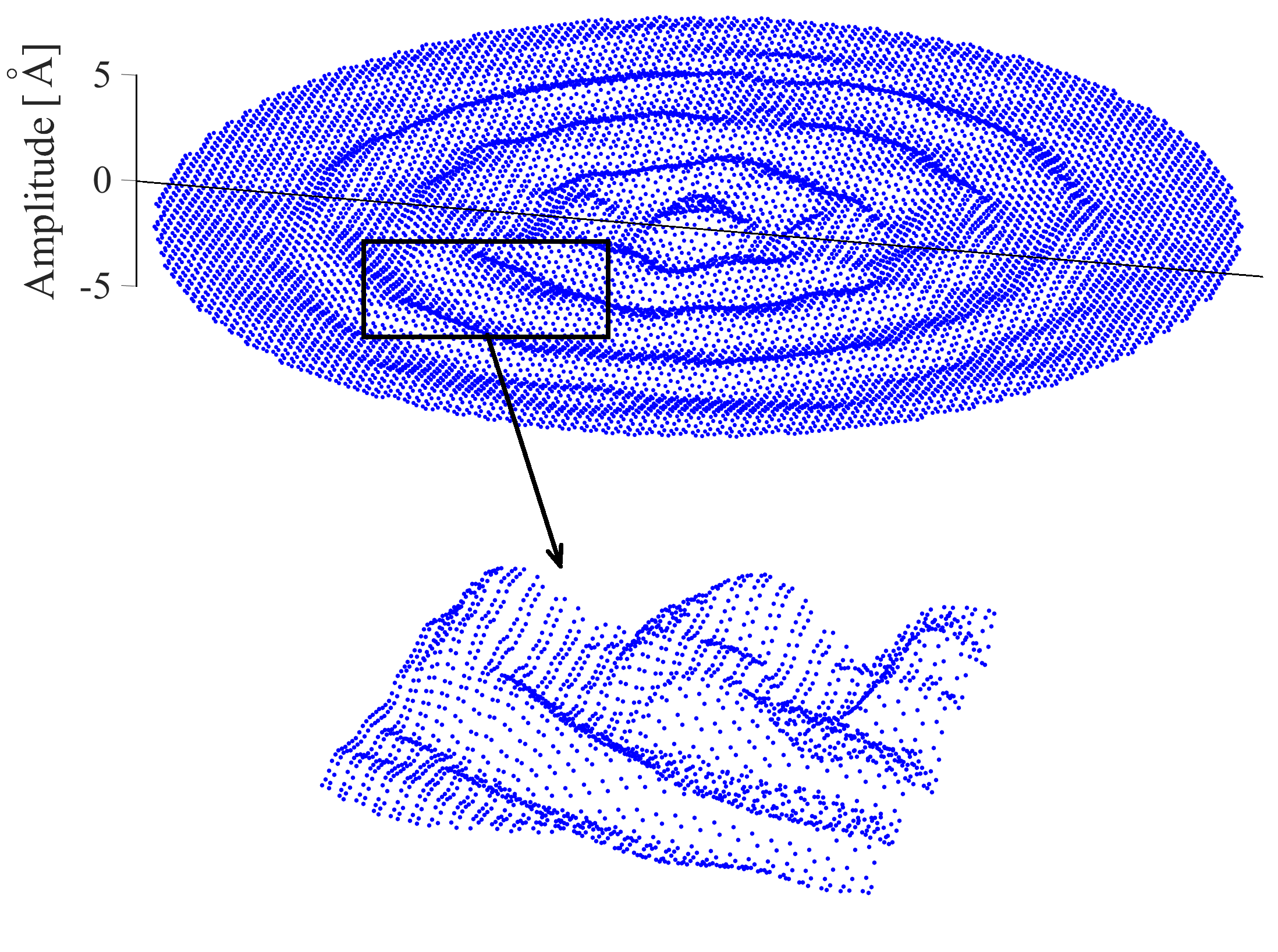}
        \caption{\label{figsnapshot}}
    \end{subfigure} 
    \caption{The schematics of MD model. \subref{figMDmodel}) The circular, flat, mono-layer graphene sheet with a radius of 10 nm (blue dots), and three rows of atoms along the boundary at which the degrees of freedom is restricted (red dots). \subref{figsnapshot}) A snapshot of the Brownian motion of the membrane with radius of $R=10$ nm, and~$T$=300~$  $K.}
    \label{figMD}
\end{figure}
\section{Numerical implementation}
\label{Numeric}
In order to perform MD simulations, we use LAMMPS software~\cite{LAMMPS}. In this software, the equations of motion are integrated using the velocity-Verlet integrator algorithm, with a time step of 1 fs. The simulations are performed for a circular, flat, mono-layer graphene sheet with a radius of 1--10 nm. The atoms in this structure are ordered in a hexagonal grid with an inter-atomic distance of 1.42 \AA  (see Figure~\ref{figMDmodel}). The edges are fully clamped by restricting the translational degrees of freedom of three rows of atoms along the boundary. The forces between atoms are described by the Tersoff potential, which is commonly used for modeling the atomic interactions in diamond, graphite, and graphene~\cite{Tersoff1988}. 

 Since the initial position of the atoms may not exactly correspond  to equilibrium or the minimum potential state, the system is relaxed by minimizing the total potential energy.  The minimization is performed by the Polak-Ribiere conjugate gradient algorithm~\cite{Polak1972}. The employed termination criteria are $1\times10^{-10}$ eV for energy or $1\times10^{-10}$ {eV/\AA}  for force. While relaxing the system, the out of plane coordinates are fixed, to prevent curling of the membrane. Next, the system is allowed to equilibrate in the constant volume and constant temperature ensemble (NVT) using the Nose-Hoover thermostat algorithm~\cite{Evans1985}. In this stage, the Nose-Hoover thermostat guarantees the Maxwell-Boltzmann velocity distribution. The algorithm is performed for 50 ps (i.e. 50000 time steps) to ensure a stable temperature is achieved. During thermalization, the boundaries of the membrane are fixed. This means the membrane will be tensioned, as a result of the negative thermal expansion of graphene. Finally, the vibration response is studied in an energy conserving ensemble (NVE). 
After the desired temperature is achieved, the thermal fluctuations of the graphene membrane are monitored for \SI{20}{\nano\second}. The atoms coordinates are saved every 0.5 ps (i.e. 500 time steps), which corresponds to approximately 20 points per vibration period of the fifth resonance of this system in 300~$  $K. Figure~\ref{figsnapshot}, shows one snapshot of the Brownian motion of a graphene membrane with a radius of 10 nm at $T$=300~$  $K. The dynamic ripples due to thermal fluctuations can be clearly observed in this figure. 

\begin{figure}[t]
        \centering
        \includegraphics[scale=0.145]{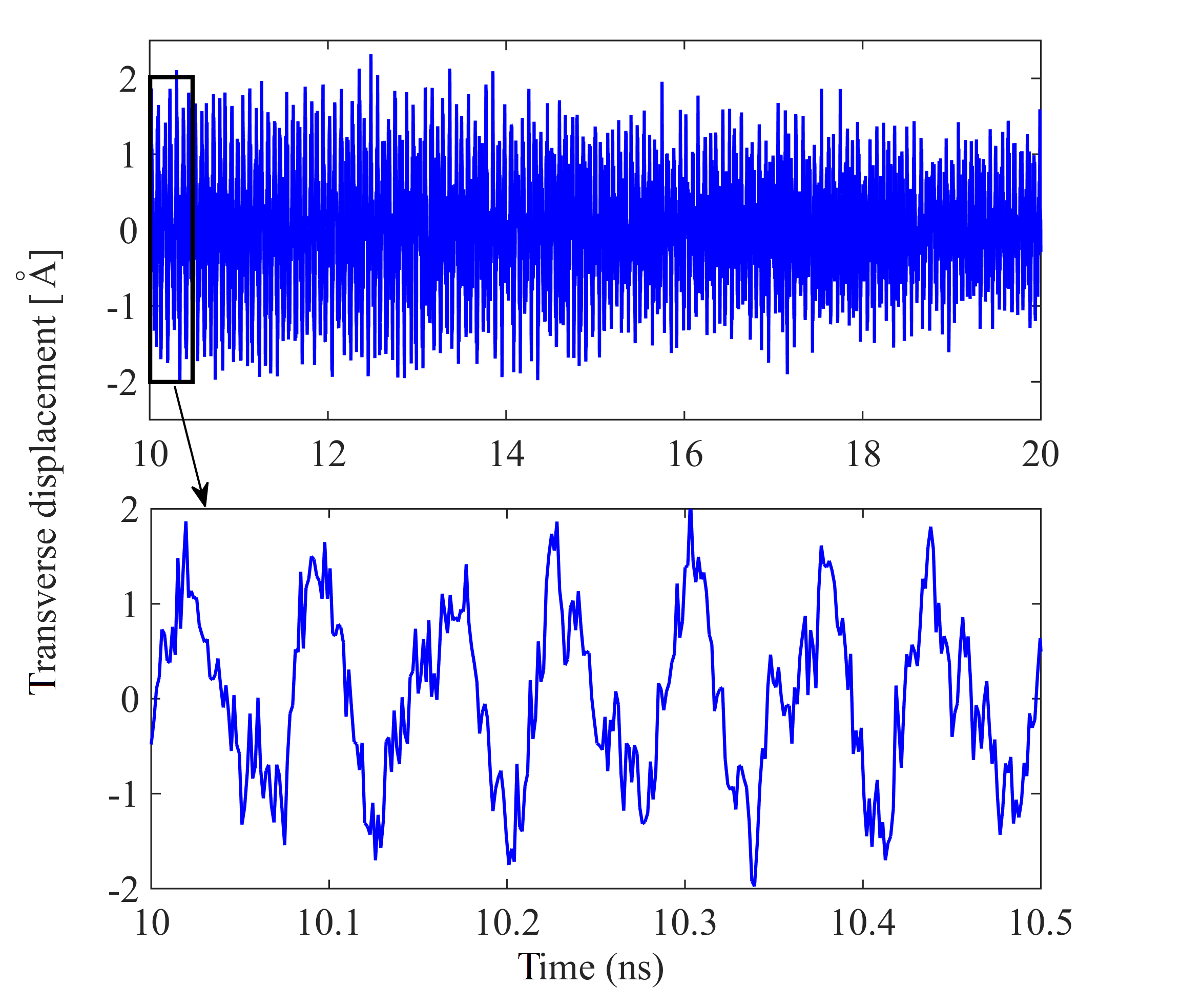}
        \caption{Transverse position of the center atom over of time, while $R=10$ nm, and~$T$=300~$  $K. }
				\label{figtimeres}
\end{figure}

The time response of the position of an atom in the center of the membrane due to these thermal fluctuations over time is  shown in Figure~\ref{figtimeres}.  It can be observed that the range of the deflection at the center of the membrane is in the order of graphene's thickness (0.335 nm). Thus, graphene at room temperature behaves as a dynamically corrugated plate that has a corrugation amplitude similar to its thickness. This shows the importance of including thermal fluctuations in estimation of graphene's mechanical properties, and also provides a mechanism by which the effective bending rigidity of graphene depends on temperature.
\begin{figure}[b]
    \centering
    \begin{subfigure}{1\linewidth}
        \centering
        \includegraphics[scale=0.15]{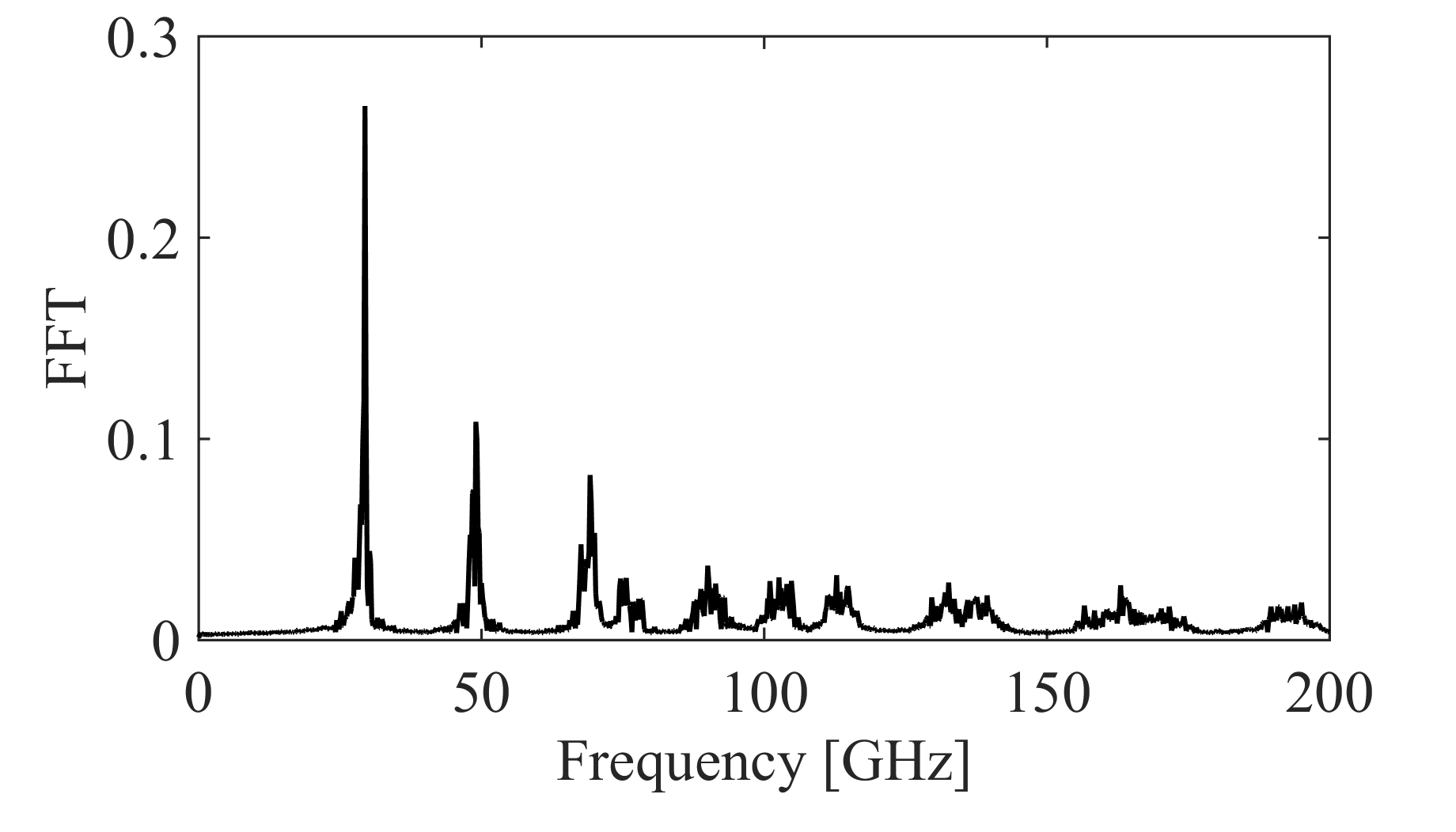}
        \caption{\label{figfft1}}
    \end{subfigure}
    \begin{subfigure}{1\linewidth}
        \centering
        \includegraphics[scale=0.15]{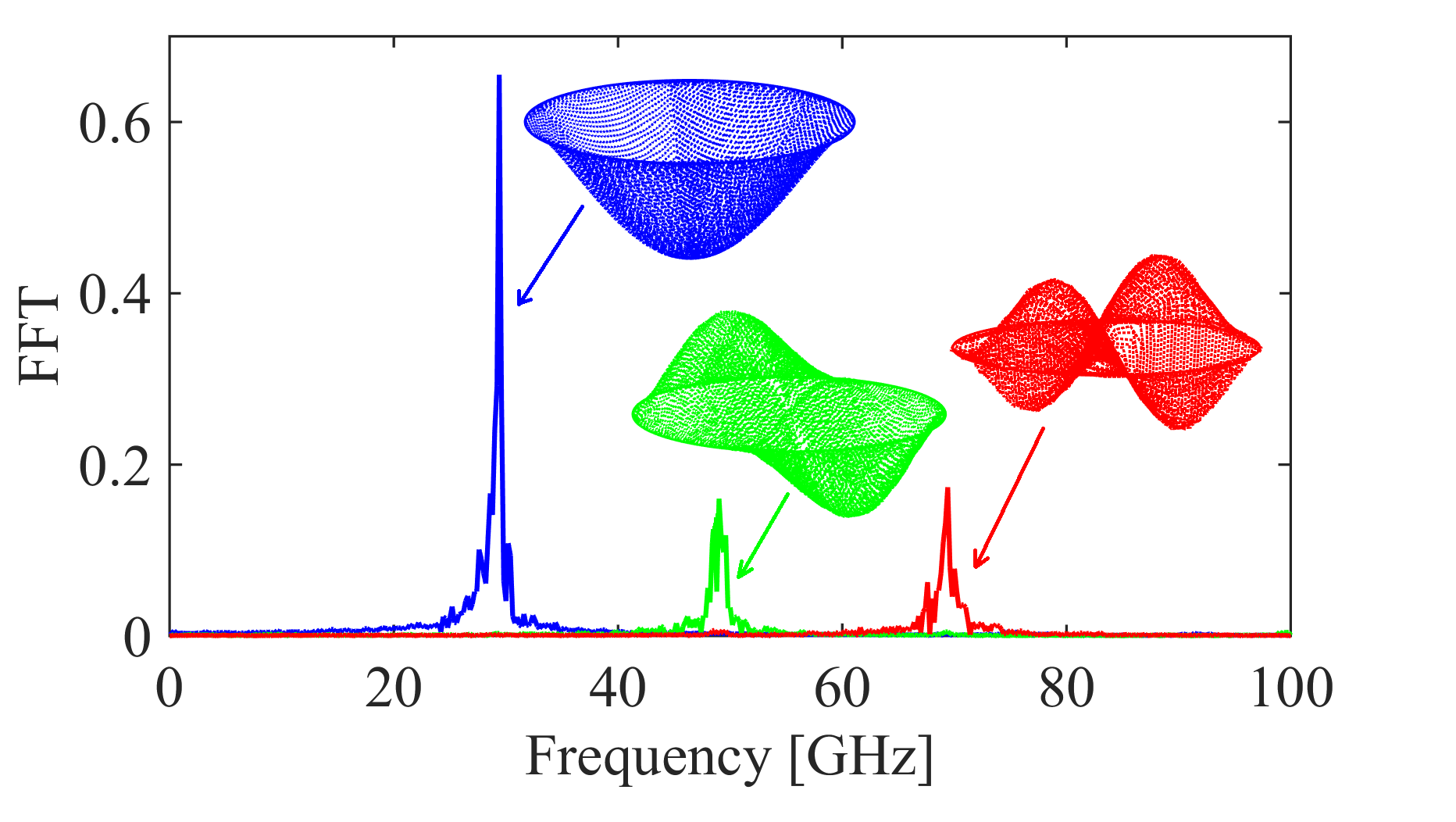}
        \caption{\label{figfft2}}
    \end{subfigure}
				\caption{ \subref{figfft1}) Averaged frequency spectrum of the time response of all atoms and \subref{figfft2}) filtered frequency spectrum for the first 3 modes, while $R=10$ nm, and~$T$=300~$  $K.}
    \label{figFFT}
\end{figure}

By applying FFT to the obtained MD time signal, the natural frequencies of the membrane are obtained. Figure~\ref{figfft1} shows the frequency spectrum obtained by averaging the FFT responses of the time signals of the atoms. 

\section{Identification technique}
\label{identification}
To identify resonance frequencies, the time response shall be filtered with respect to the associated modes. 
 This filtering is performed by using the orthogonality of vibration modes, i.e. by projecting the time response on a certain mode shape~\cite{Rao2007}. This projection shall be performed via a dot product between the snap-shots of the MD transverse motion and the vector describing the vibration modes at the position of all atoms. The analytic solutions for the mode shapes of a circular clamped membrane are used for the  vibration modes \cite{SM}. For each of the mode shapes, a time-trace of the resulting dot product is determined and an FFT is applied. Figure \ref{figfft2} shows the filtered frequency response of the first few modes of vibrations, indicated in different colors. By determining the peak frequency of each of the mode shapes, the first 10 resonance frequencies (i.e. $\omega_i^{MD}$) of the MD model are determined.

\begin{figure*}
    \centering
    \begin{subfigure}[b]{0.3\linewidth}
        \centering
        \includegraphics[scale=0.148]{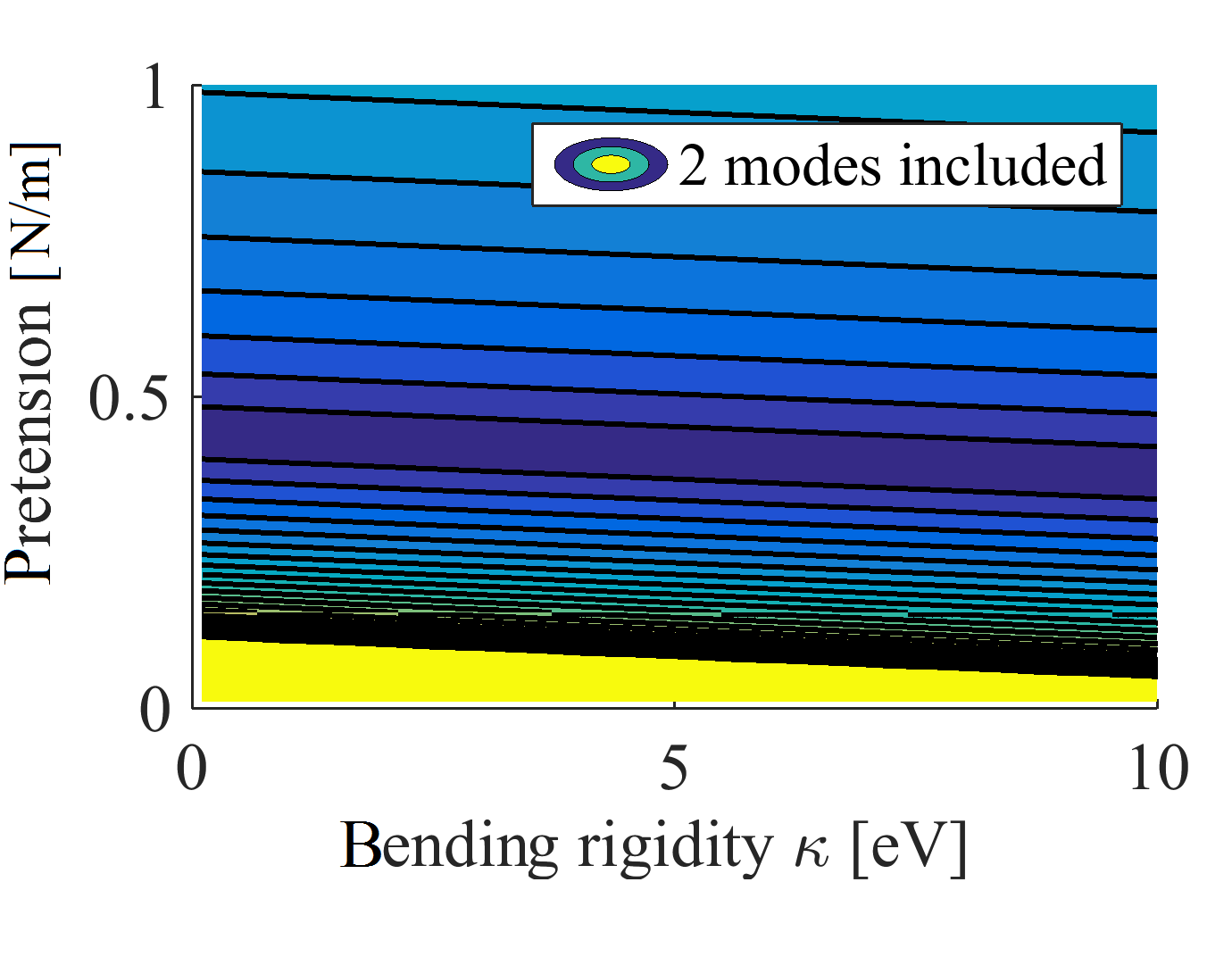}
        \caption{\label{figerrorsurf2}}
    \end{subfigure}
    \begin{subfigure}[b]{0.3\linewidth}
        \includegraphics[scale=0.148]{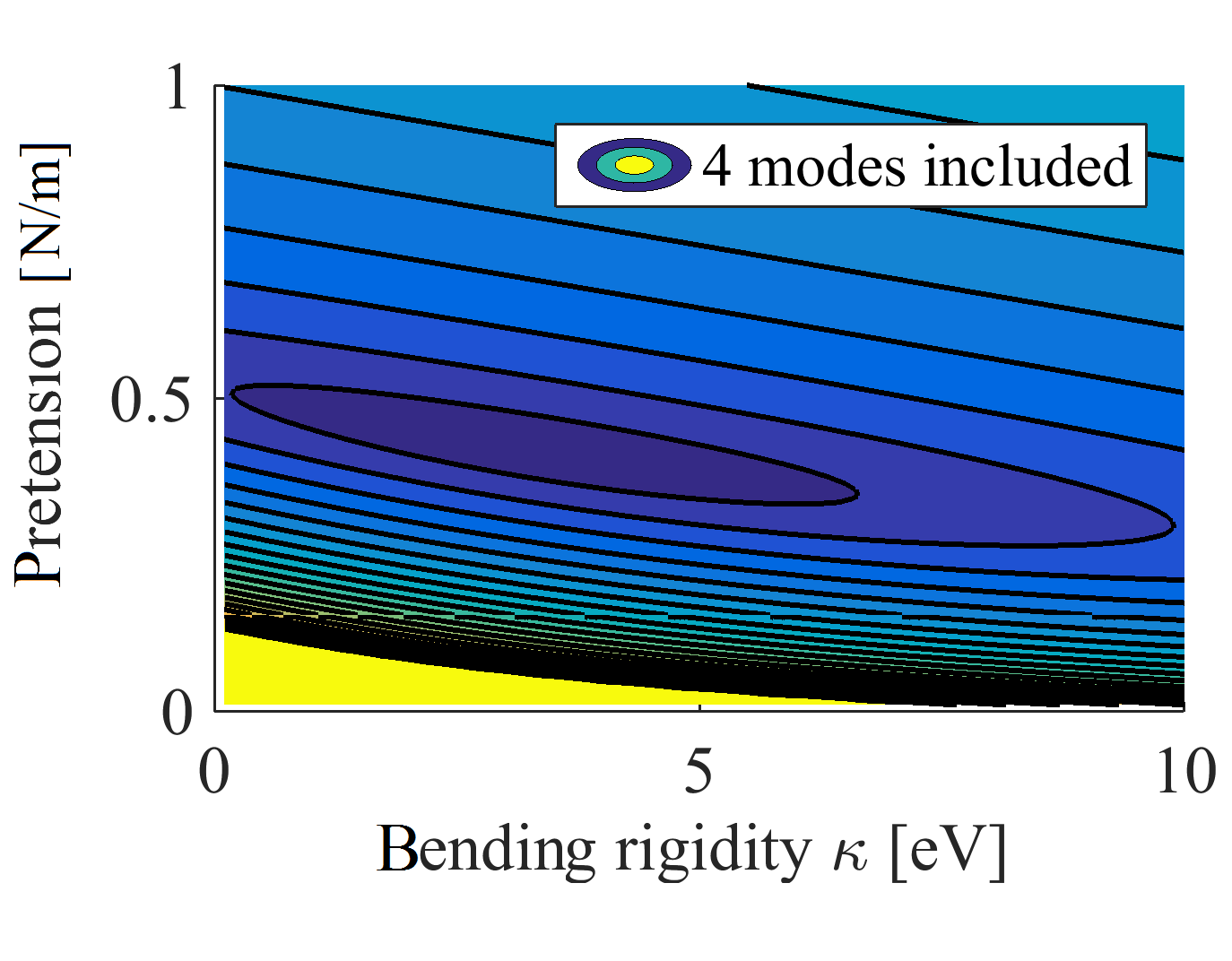}
        \caption{\label{figerrorsurf4}}
    \end{subfigure}
    \begin{subfigure}[b]{0.3\linewidth}
        \includegraphics[scale=0.148]{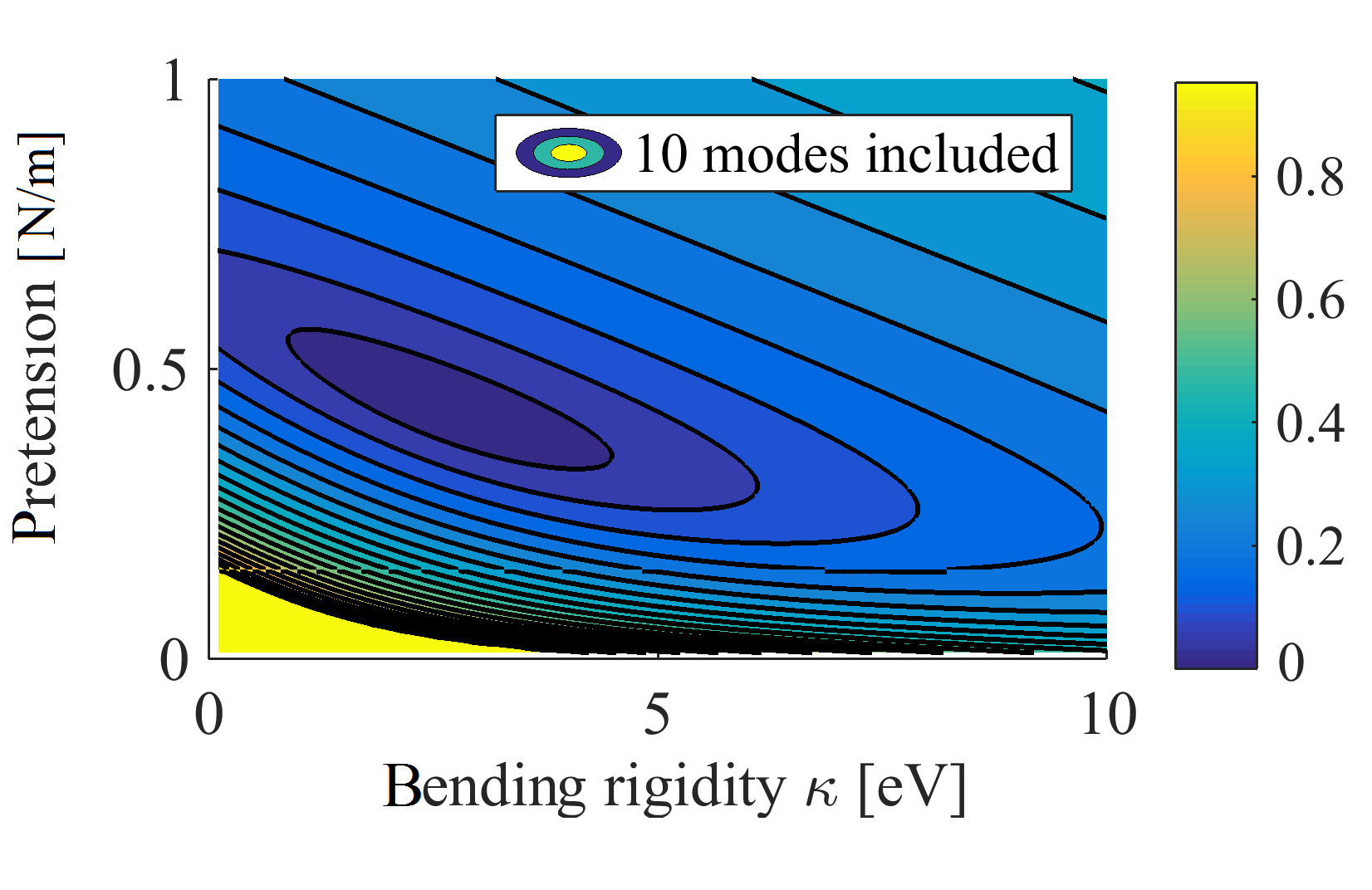}
        \caption{\label{figerrorsurf10}}
    \end{subfigure}
    \caption{ The normalized error ($e$) as a function of the fitting parameters $\kappa$ and $n_0$, when including \subref{figerrorsurf2})~$N=2$, \subref{figerrorsurf4})~$N=4$ , and \subref{figerrorsurf10})~$N=10$ frequencies, while $R=10$ nm, and~$T$=300$   $K.  }
    \label{figerrorsurf}
\end{figure*}

\begin{figure*}[t!]
    \centering
      \begin{subfigure}[t]{0.45\textwidth}
    \includegraphics[scale=0.155]{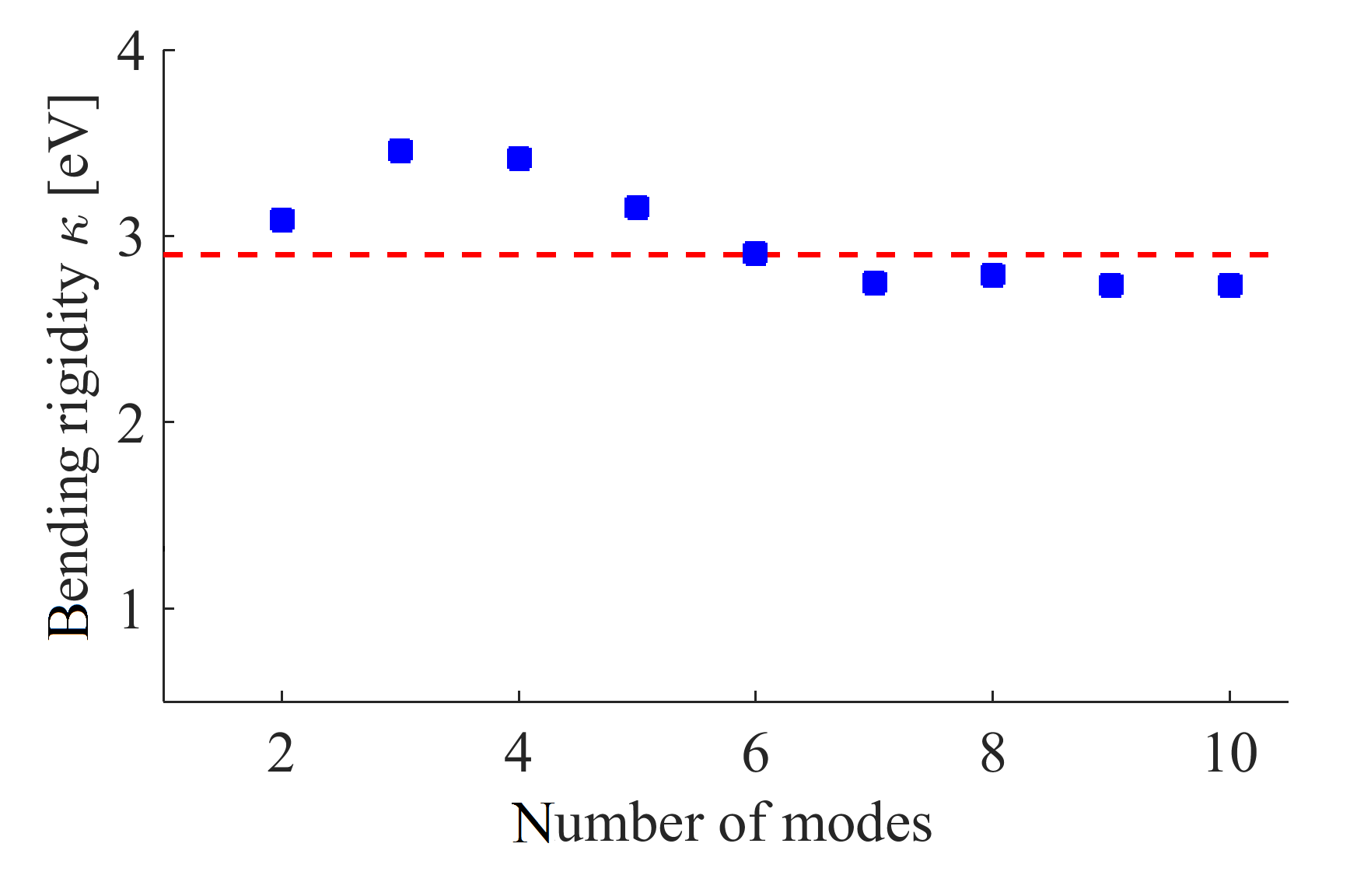}
    \caption{\label{figkmodes}}
\end{subfigure}
 \begin{subfigure}[t]{0.45\textwidth}
 \centering
    \includegraphics[scale=0.155]{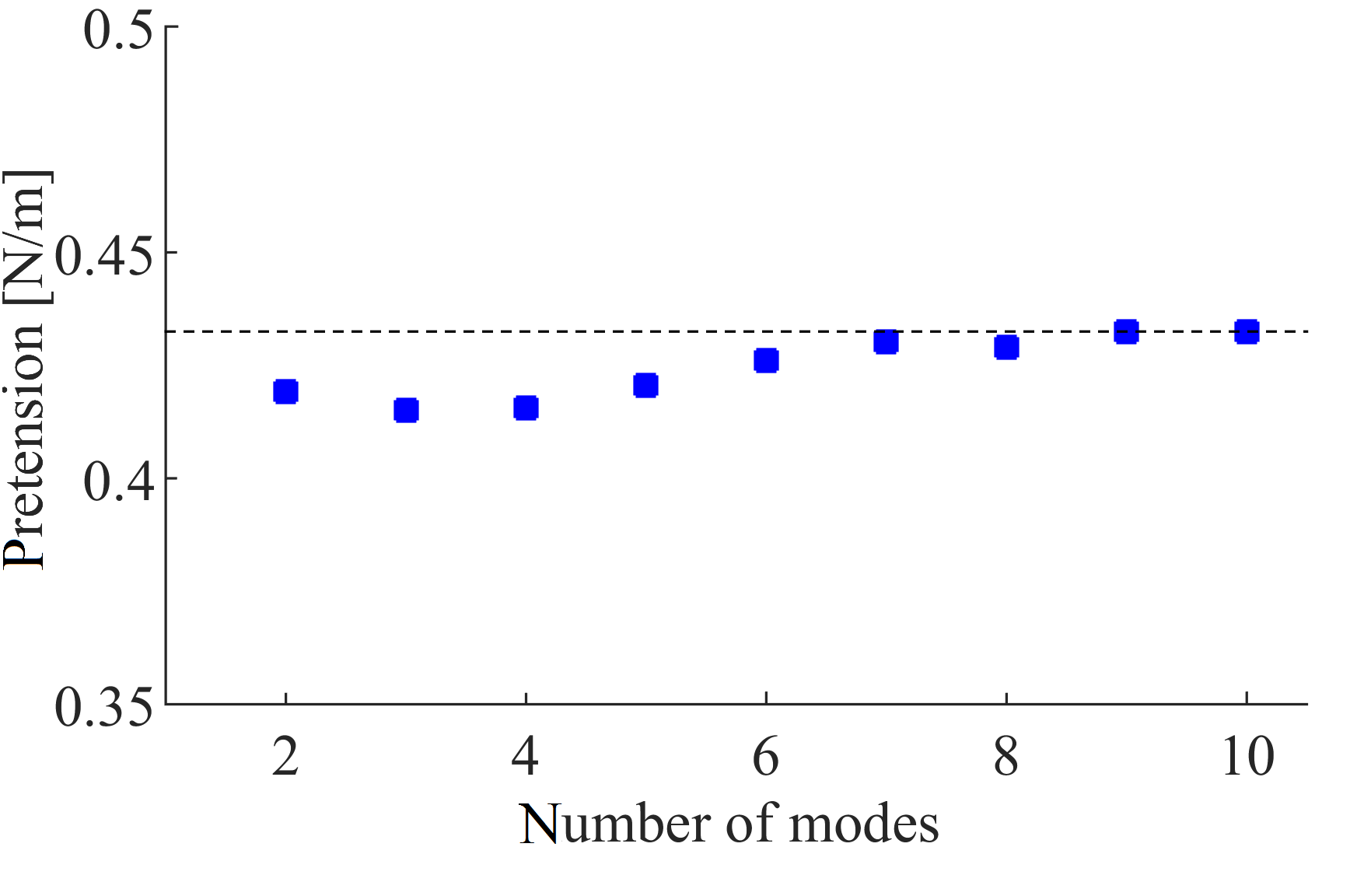}
    \caption{\label{fign0mode}}
    \end{subfigure}
    \caption{a) The obtained bending rigidity $\kappa$ as a function of the number of frequencies in the fitting process (blue dots) and the approximated one with  $q_{\mathrm{eff}}$=$\pi/R$ (red dashed line) from Equation~(\ref{bendingkatsnelson}) \cite{Katsnelson2011}, for $R=10$ nm at~$T$=300~$  $K. b) The obtained pretension $n_0$ as a function of the number of frequencies in the fitting process (blue dots) converging to a pretension of \SI{0.43}{\newton/\meter} (black dashed line), for $R=10$ nm at~$T$=300 K.}
\end{figure*}

Next, we obtain the equations of motion by using the von K\'arman plate theory~\cite{Amabili2008} and by following Lagrangian approach. For more details about our continuum mechanics model, see the online Supplemental Material \cite{SM}. 
In our formulation, bending rigidity ($\kappa$) and the pretension ($n_0$) of the membrane are considered to be  unknown parameters that will be calibrated by means of MD simulations. By using the proposed approach,  a set of $N$  equations describing the motion of the membrane are obtained as follows:
\begin{equation}
\label{eqCMeom}
\mathbf{M} \mathbf{\ddot{q} }+ \mathbf{K}\mathbf{{q} } = \mathbf{0} ,
\end{equation}
where $\mathbf{M}$ and $\mathbf{K}$ are the equivalent mass and stiffness matrices, respectively. Moreover, $\mathbf{q}$ is the vector comprising of the $N$ time dependent generalized coordinates defining the motion of the membrane. The resonance frequencies can be directly determined from the characteristic equation of this system, i.e. $\det \left( \mathbf{M}^{-1} \mathbf{K} - \mathbf{I} (\omega^2 \right) = \mathbf{0}$. It is worth noting that the stiffness matrix ($\mathbf{K}$), and hence the obtained resonance frequencies $\omega_i^{CM}$ will be functions of the pretension ($n_0$) and bending rigidity ($\kappa$).  
Moreover, the obtained frequencies are independent of the value of the elastic modulus, since in the  continuum framework modeling of membranes, the elastic modulus only affects the nonlinear dynamics of the membrane at large amplitudes~\cite{Farbod2017}, and not the linear response.

Next, the resonance frequencies from CM (i.e. $\omega_i^{CM}(\kappa, n_0)$) is numerically fitted to the obtained set of resonance frequencies from MD (i.e. $\omega_i^{MD}$). The fitting is performed by a least squares method and using  $\kappa$ and $n_0$ as fit parameters. The squared normalized error of $N$ resonance frequencies between the two methods is minimized, where the error is defined as:
\begin{equation}
e=\sqrt{\frac{\sum_{i=1}^{N}\big({\frac{\omega_i^{MD}-\omega_i^{CM}(\kappa, n_0)}{\omega_i^{CM}}}\big)^2}{N}},
\label{error}
\end{equation}
It shall be noted that mathematically, only two resonance frequencies are needed to determine $\kappa$ and $n_0$, since it involves solving 2 equations with 2 unknowns.  However, retaining higher modes is necessary to increase the accuracy because the radius of curvature of the membrane at higher frequency modes is relatively smaller, and therefore, the associated resonance frequencies are more sensitive to the bending rigidity. Moreover, by employing a higher number of degrees of freedom, one can assure that the model in (\ref{eqCMeom}) can better describe the dynamic ripples due to Brownian motion.
 
The error between the natural frequencies obtained via CM and MD models decreases by including higher modes in the fitting process and leads to a converged value for the bending rigidity. Figure~\ref{figerrorsurf} shows the normalized error ($e$) obtained from Equation (3), as a function of the fitting parameters.  This figure confirms that including higher modes in the fitting process decreases the surface area of the minimum error, and leads to a more accurate bending rigidity.  These graphs clearly show the necessity of incorporating multiple modes in the approximation in order to reach a converged solution.

\section{Results and discussion}
\label{results}
The convergence of the bending rigidity and pretension versus the number of modes retained in the fitting procedure is shown in Figure~\ref{figkmodes} and Figure~\ref{fign0mode}, respectively. It is seen that, at room temperature, by including 10 natural frequencies, the solution converges to a bending rigidity of \SI{2.7}{\electronvolt} and the corresponding pretension due to thermal strain is obtained as \SI{0.43}{\newton/\meter}. For $E_{2D}=340$~N/m and Poisson's ratio $\nu=0.17$, this value corresponds to a thermal expansion coefficient of $-3.52\times 10^{-6}$ 1/K at room temperature (see Equation~(4) in the Supplemental Material \cite{SM}) and is in agreement with first principle calculations \cite{mounet2005first}. 

Moreover, it can be observed from Figure~\ref{figkmodes} that the obtained effective bending rigidity is converging to the bending rigidity obtained by Equation~(\ref{bendingkatsnelson})~\cite{Katsnelson2011} when using an effective wave number $q_{\mathrm{eff}}=\pi/R$. It should be noted that $q_{\mathrm{eff}}$ is found between discrete wave numbers that fit in the membrane.
The obtained value of $\kappa$ and $q_{\mathrm{eff}}$ are not only affected by the simultaneous fit of 10 modes with different wavelengths, but also depend on circular geometry of the drum. 

\begin{figure}[t!]
    \centering
    \includegraphics[scale=0.155]{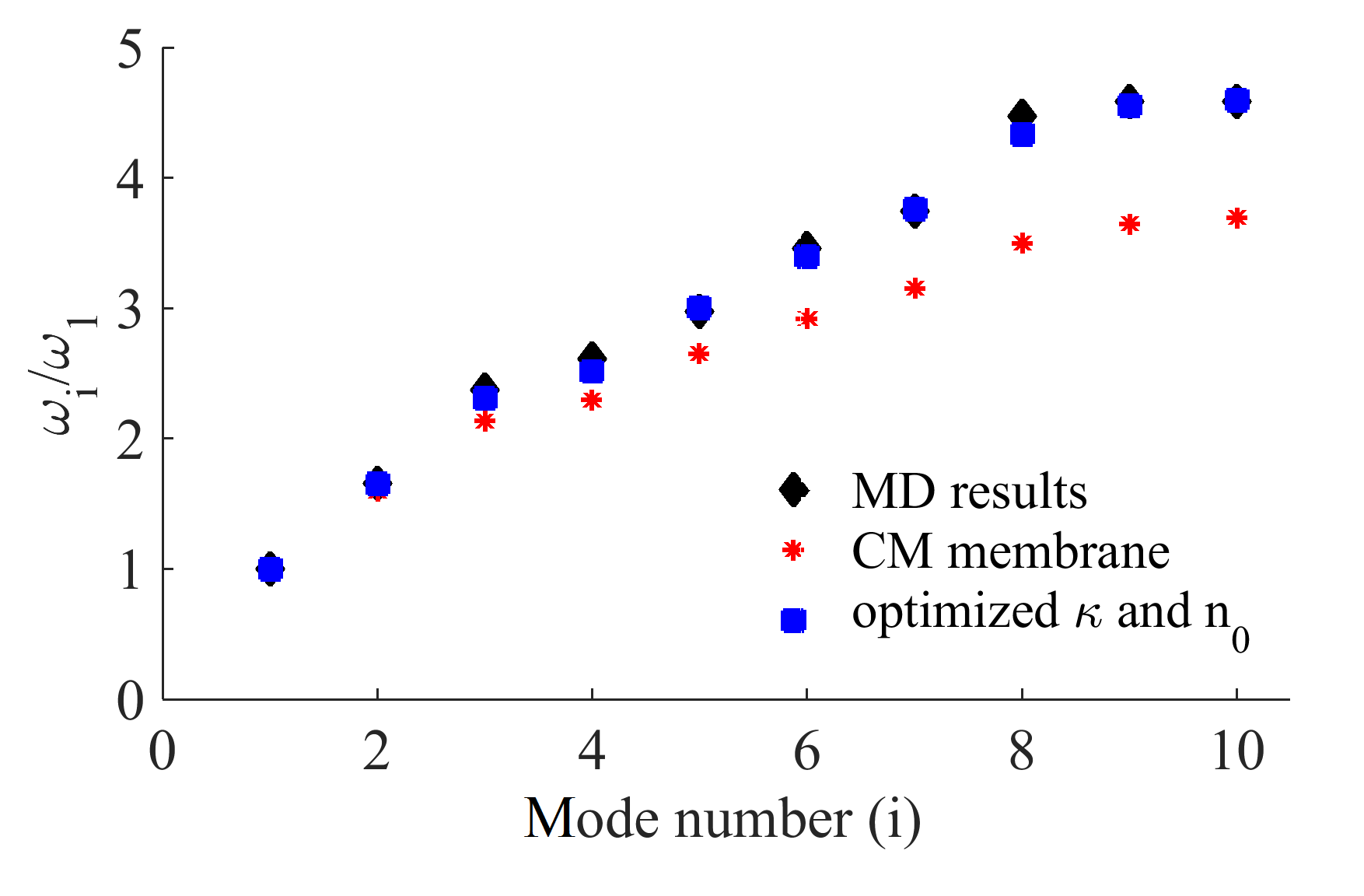}
    \caption{The normalized natural frequencies versus the mode number, obtained from MD, the proposed model with the optimized parameters ($n_0=\SI{0.43}{\newton/\meter}=$ and $\kappa=\SI{2.7}{\electronvolt}$), and classical membrane theory where $\kappa=0$, for $R=10$ nm at~$T$=300  K. }
    \label{figfreqmode}
\end{figure}

The ratio between the first 10 resonance frequencies and the fundamental frequency ($\omega_1=\SI{28.8}{\giga\hertz}$) are shown in Figure~\ref{figfreqmode}. For comparison, the results of MD simulations and those obtained from a classical membrane model (with zero bending rigidity) are also plotted in this figure. As can be observed, by using a single optimized value for pretension and bending rigidity, our CM model can very well reproduce all the 10 natural frequencies of the MD model, while it is clear that a membrane model that neglects the bending rigidity of graphene cannot capture the observed dynamic behavior, especially for the higher resonance modes. 

\begin{figure}[t!]
    \centering
    \includegraphics[scale=0.155]{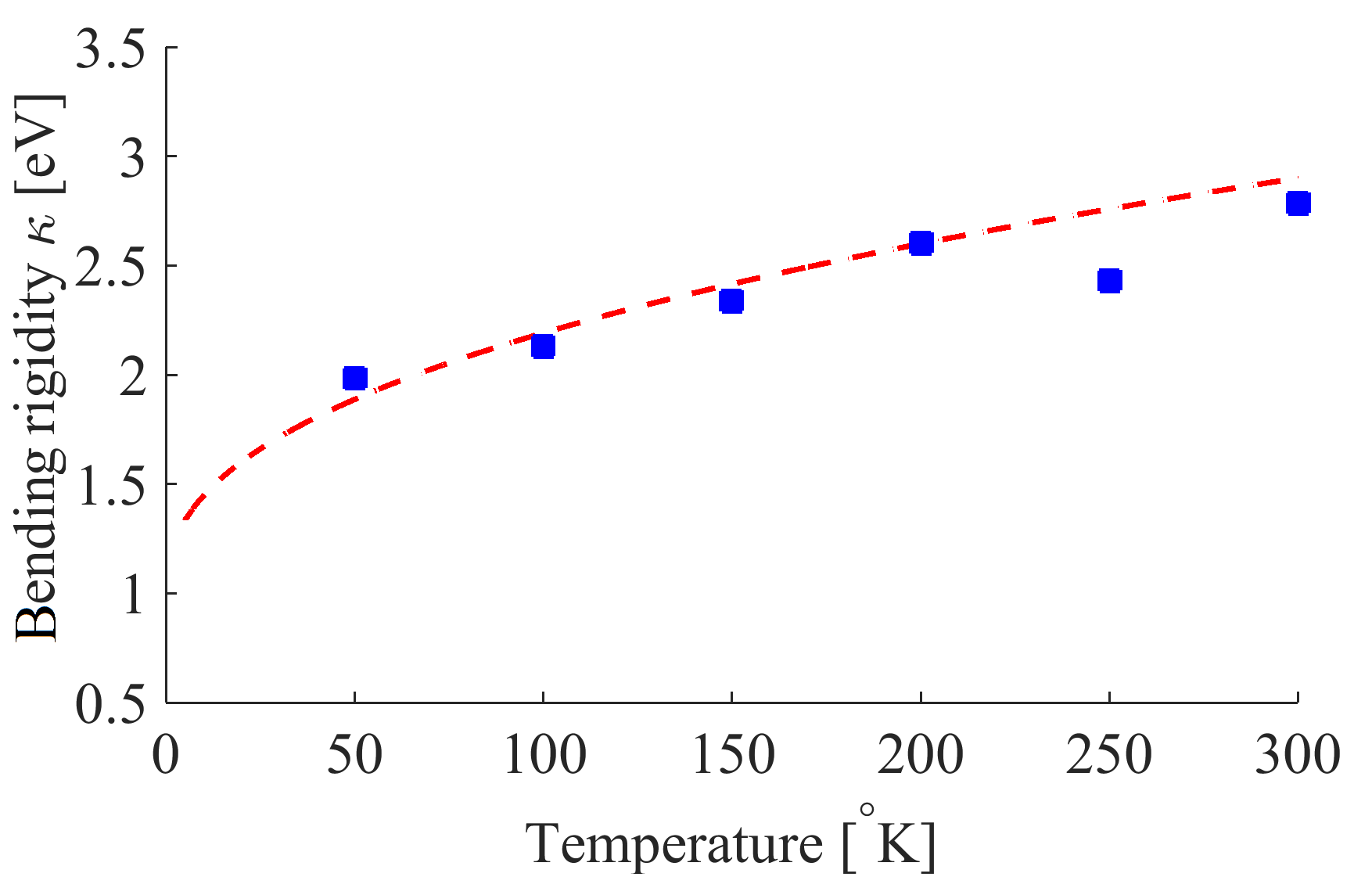}
        \caption{The obtained bending rigidity $\kappa$  (blue dots), and the approximated one with  $q_{\mathrm{eff}}=\pi/R$(red dashed line) from Equation~(\ref{bendingkatsnelson}) \cite{Katsnelson2011}, as a function of the temperature, for~$R=10$~nm.}
    \label{figkT}
\end{figure}
Furthermore, using the proposed method, the temperature, and size dependence of the bending rigidity can be studied. In this regard, Figure~\ref{figkT} shows the obtained bending rigidity as a function of temperature. Included in the Figure is also the bending rigidity obtained from Equation~(\ref{bendingkatsnelson}) with $q_{\mathrm{eff}}$=$\pi/R$. 
As can be seen, both methods predict an increase in the bending rigidity with increasing temperature. This increase is due to  entropic effects in graphene. In fact, graphene's bending rigidity resembles an entropic spring, like a rubber band, in which entropy and thermodynamics affect elasticity. In such systems, the free energy $A=U-TS$ is a sum of the internal energy $U$ and the product of temperature $T$ and entropy $S$. The external force $F$ needed for reversible isothermal extension of such a spring is $F=dA/dx=dU/dx-TdS/dx=k(T)x$. Therefore, the effective stiffness $k(T)$ increases with temperature due to the reduction in entropy ($dS/dx<0$) upon elongation  in the spring or rubber band. 

 \begin{figure}[t]
    \centering
    \includegraphics[scale=0.155]{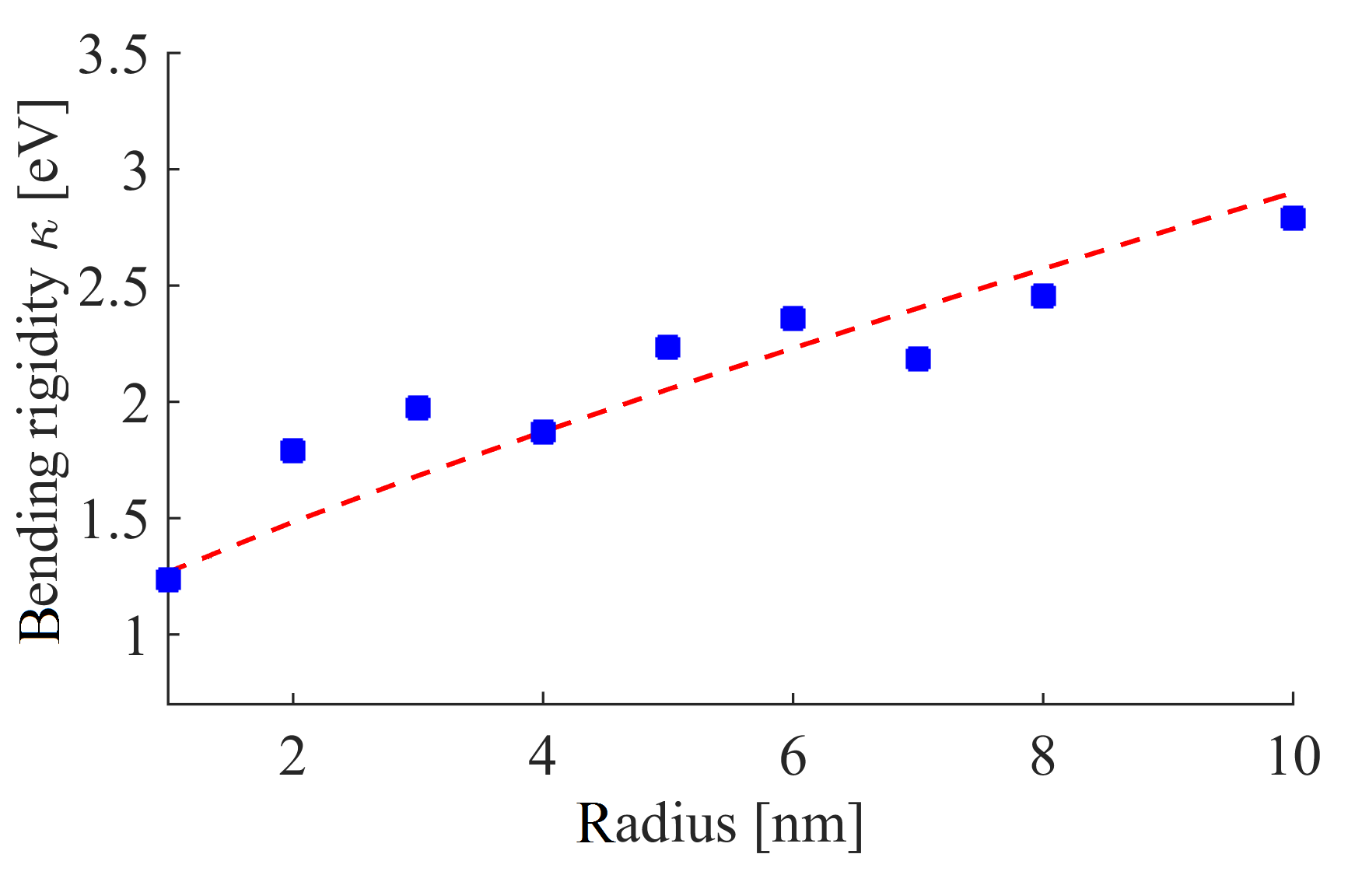}
    \caption{The obtained bending rigidity $\kappa$  (blue dots) and the approximated one with  $q_{\mathrm{eff}}$=$\pi/R$(red dashed line) from Equation~(\ref{bendingkatsnelson}) \cite{Katsnelson2011}, at~$T$=300~$  $K. }
    \label{figkR}
\end{figure}
In Figure~\ref{figkR}, we report the bending rigidity for different radii of the membrane. It can be seen that the bending rigidity increases monotonically with the radius of the membrane, and it fits Equation~(\ref{bendingkatsnelson}) when $q_{\mathrm{eff}}=\pi/R$. With this value of $q_{\mathrm{eff}}$, for a membrane with $R=\SI{5}{\micro\meter}$, Equation~(\ref{bendingkatsnelson}) suggests a re-normalized bending rigidity of 375 eV. 
This size dependence can be attributed to two main reasons: (i) at small scales the atoms are more bounded for free thermal fluctuations and therefore, they appear as relatively less dynamic as compared to larger scales; and (ii) at small scales the macroscopic and microscopic bending rigidities are physically non-distinguishable. As a result, our obtained bending rigidity at $R=1$ nm is close to the microscopic temperature-independent values of \SI{1.4}{\electronvolt}--\SI{1.6}{\electronvolt}~\cite{lu2009,  Koskinen2010,lu2009,Nicholl2015,Kudin2001,Sanchez1999}.

\section{Conclusions}
In conclusion, we used modal analysis for direct estimation of the macroscopic bending rigidity of graphene membranes at high frequencies. The current work confirms that the bending rigidity in graphene membranes depends on the temperature and membrane size. In particular, the size-dependence of the bending rigidity is a special property that is not encountered in macroscopic systems. Moreover, our obtained bending rigidity is  in agreement with the size-dependent renormalized bending rigidity predicted by the statistical mechanics of elastic membranes.
Our method is not only suitable for obtaining the bending rigidity of graphene but is also useful for characterization of other nano-materials at high frequencies, while incorporating thermal fluctuations. \\

\section*{Acknowledgements}
We acknowledge productive discussions with Y.M. Blanter from TU Delft and M. Katsnelson from Radboud University, The Netherlands. BS and FA further acknowledge the financial support from TU Delft, 3mE cohesion grant NITRO.



\begin{thebibliography}{36}%
\makeatletter
\providecommand \@ifxundefined [1]{%
 \@ifx{#1\undefined}
}%
\providecommand \@ifnum [1]{%
 \ifnum #1\expandafter \@firstoftwo
 \else \expandafter \@secondoftwo
 \fi
}%
\providecommand \@ifx [1]{%
 \ifx #1\expandafter \@firstoftwo
 \else \expandafter \@secondoftwo
 \fi
}%
\providecommand \natexlab [1]{#1}%
\providecommand \enquote  [1]{``#1''}%
\providecommand \bibnamefont  [1]{#1}%
\providecommand \bibfnamefont [1]{#1}%
\providecommand \citenamefont [1]{#1}%
\providecommand \href@noop [0]{\@secondoftwo}%
\providecommand \href [0]{\begingroup \@sanitize@url \@href}%
\providecommand \@href[1]{\@@startlink{#1}\@@href}%
\providecommand \@@href[1]{\endgroup#1\@@endlink}%
\providecommand \@sanitize@url [0]{\catcode `\\12\catcode `\$12\catcode
  `\&12\catcode `\#12\catcode `\^12\catcode `\_12\catcode `\%12\relax}%
\providecommand \@@startlink[1]{}%
\providecommand \@@endlink[0]{}%
\providecommand \url  [0]{\begingroup\@sanitize@url \@url }%
\providecommand \@url [1]{\endgroup\@href {#1}{\urlprefix }}%
\providecommand \urlprefix  [0]{URL }%
\providecommand \Eprint [0]{\href }%
\providecommand \doibase [0]{http://dx.doi.org/}%
\providecommand \selectlanguage [0]{\@gobble}%
\providecommand \bibinfo  [0]{\@secondoftwo}%
\providecommand \bibfield  [0]{\@secondoftwo}%
\providecommand \translation [1]{[#1]}%
\providecommand \BibitemOpen [0]{}%
\providecommand \bibitemStop [0]{}%
\providecommand \bibitemNoStop [0]{.\EOS\space}%
\providecommand \EOS [0]{\spacefactor3000\relax}%
\providecommand \BibitemShut  [1]{\csname bibitem#1\endcsname}%
\let\auto@bib@innerbib\@empty
\bibitem [{\citenamefont {Dolleman}\ \emph {et~al.}(2015)\citenamefont
  {Dolleman}, \citenamefont {Davidovikj}, \citenamefont {Cartamil-Bueno},
  \citenamefont {van~der Zant},\ and\ \citenamefont
  {Steeneken}}]{Dolleman2015}%
  \BibitemOpen
  \bibfield  {author} {\bibinfo {author} {\bibfnamefont {R.~J.}\ \bibnamefont
  {Dolleman}}, \bibinfo {author} {\bibfnamefont {D.}~\bibnamefont
  {Davidovikj}}, \bibinfo {author} {\bibfnamefont {S.~J.}\ \bibnamefont
  {Cartamil-Bueno}}, \bibinfo {author} {\bibfnamefont {H.~S.}\ \bibnamefont
  {van~der Zant}}, \ and\ \bibinfo {author} {\bibfnamefont {P.~G.}\
  \bibnamefont {Steeneken}},\ }\href@noop {} {\bibfield  {journal} {\bibinfo
  {journal} {Nano letters}\ }\textbf {\bibinfo {volume} {16}},\ \bibinfo
  {pages} {568} (\bibinfo {year} {2015})}\BibitemShut {NoStop}%
\bibitem [{\citenamefont {Atalaya}\ \emph {et~al.}(2010)\citenamefont
  {Atalaya}, \citenamefont {Kinaret},\ and\ \citenamefont
  {Isacsson}}]{Atalaya2010}%
  \BibitemOpen
  \bibfield  {author} {\bibinfo {author} {\bibfnamefont {J.}~\bibnamefont
  {Atalaya}}, \bibinfo {author} {\bibfnamefont {J.~M.}\ \bibnamefont
  {Kinaret}}, \ and\ \bibinfo {author} {\bibfnamefont {A.}~\bibnamefont
  {Isacsson}},\ }\href@noop {} {\bibfield  {journal} {\bibinfo  {journal} {EPL
  (Europhysics Letters)}\ }\textbf {\bibinfo {volume} {91}},\ \bibinfo {pages}
  {48001} (\bibinfo {year} {2010})}\BibitemShut {NoStop}%
\bibitem [{\citenamefont {Schedin}\ \emph {et~al.}(2007)\citenamefont
  {Schedin}, \citenamefont {Geim}, \citenamefont {Morozov}, \citenamefont
  {Hill}, \citenamefont {Blake}, \citenamefont {Katsnelson},\ and\
  \citenamefont {Novoselov}}]{Schedin2007}%
  \BibitemOpen
  \bibfield  {author} {\bibinfo {author} {\bibfnamefont {F.}~\bibnamefont
  {Schedin}}, \bibinfo {author} {\bibfnamefont {A.}~\bibnamefont {Geim}},
  \bibinfo {author} {\bibfnamefont {S.}~\bibnamefont {Morozov}}, \bibinfo
  {author} {\bibfnamefont {E.}~\bibnamefont {Hill}}, \bibinfo {author}
  {\bibfnamefont {P.}~\bibnamefont {Blake}}, \bibinfo {author} {\bibfnamefont
  {M.}~\bibnamefont {Katsnelson}}, \ and\ \bibinfo {author} {\bibfnamefont
  {K.}~\bibnamefont {Novoselov}},\ }\href@noop {} {\bibfield  {journal}
  {\bibinfo  {journal} {Nature materials}\ }\textbf {\bibinfo {volume} {6}},\
  \bibinfo {pages} {652} (\bibinfo {year} {2007})}\BibitemShut {NoStop}%
\bibitem [{\citenamefont {Schwierz}(2010)}]{Schwierz2010}%
  \BibitemOpen
  \bibfield  {author} {\bibinfo {author} {\bibfnamefont {F.}~\bibnamefont
  {Schwierz}},\ }\href@noop {} {\bibfield  {journal} {\bibinfo  {journal}
  {Nature nanotechnology}\ }\textbf {\bibinfo {volume} {5}},\ \bibinfo {pages}
  {487} (\bibinfo {year} {2010})}\BibitemShut {NoStop}%
\bibitem [{\citenamefont {Chen}\ \emph {et~al.}(2013)\citenamefont {Chen},
  \citenamefont {Lee}, \citenamefont {Deshpande}, \citenamefont {Lee},
  \citenamefont {Lekas}, \citenamefont {Shepard},\ and\ \citenamefont
  {Hone}}]{Changyao2013}%
  \BibitemOpen
  \bibfield  {author} {\bibinfo {author} {\bibfnamefont {C.}~\bibnamefont
  {Chen}}, \bibinfo {author} {\bibfnamefont {S.}~\bibnamefont {Lee}}, \bibinfo
  {author} {\bibfnamefont {V.~V.}\ \bibnamefont {Deshpande}}, \bibinfo {author}
  {\bibfnamefont {G.-H.}\ \bibnamefont {Lee}}, \bibinfo {author} {\bibfnamefont
  {M.}~\bibnamefont {Lekas}}, \bibinfo {author} {\bibfnamefont
  {K.}~\bibnamefont {Shepard}}, \ and\ \bibinfo {author} {\bibfnamefont
  {J.}~\bibnamefont {Hone}},\ }\href {\doibase 10.1038/nnano.2013.232}
  {\bibfield  {journal} {\bibinfo  {journal} {Nature nanotechnology}\ }\textbf
  {\bibinfo {volume} {8}},\ \bibinfo {pages} {923} (\bibinfo {year}
  {2013})}\BibitemShut {NoStop}%
\bibitem [{\citenamefont {Novoselov}\ \emph {et~al.}(2012)\citenamefont
  {Novoselov}, \citenamefont {Fal}, \citenamefont {Colombo}, \citenamefont
  {Gellert}, \citenamefont {Schwab},\ and\ \citenamefont
  {Kim}}]{Novoselov2012}%
  \BibitemOpen
  \bibfield  {author} {\bibinfo {author} {\bibfnamefont {K.~S.}\ \bibnamefont
  {Novoselov}}, \bibinfo {author} {\bibfnamefont {V.}~\bibnamefont {Fal}},
  \bibinfo {author} {\bibfnamefont {L.}~\bibnamefont {Colombo}}, \bibinfo
  {author} {\bibfnamefont {P.}~\bibnamefont {Gellert}}, \bibinfo {author}
  {\bibfnamefont {M.}~\bibnamefont {Schwab}}, \ and\ \bibinfo {author}
  {\bibfnamefont {K.}~\bibnamefont {Kim}},\ }\href@noop {} {\bibfield
  {journal} {\bibinfo  {journal} {Nature}\ }\textbf {\bibinfo {volume} {490}},\
  \bibinfo {pages} {192} (\bibinfo {year} {2012})}\BibitemShut {NoStop}%
\bibitem [{\citenamefont {Zhao}\ \emph {et~al.}(2009)\citenamefont {Zhao},
  \citenamefont {Min},\ and\ \citenamefont {Aluru}}]{Zhao2009}%
  \BibitemOpen
  \bibfield  {author} {\bibinfo {author} {\bibfnamefont {H.}~\bibnamefont
  {Zhao}}, \bibinfo {author} {\bibfnamefont {K.}~\bibnamefont {Min}}, \ and\
  \bibinfo {author} {\bibfnamefont {N.}~\bibnamefont {Aluru}},\ }\href@noop {}
  {\bibfield  {journal} {\bibinfo  {journal} {Nano letters}\ }\textbf {\bibinfo
  {volume} {9}},\ \bibinfo {pages} {3012} (\bibinfo {year} {2009})}\BibitemShut
  {NoStop}%
\bibitem [{\citenamefont {Eckmann}\ \emph {et~al.}(2012)\citenamefont
  {Eckmann}, \citenamefont {Felten}, \citenamefont {Mishchenko}, \citenamefont
  {Britnell}, \citenamefont {Krupke}, \citenamefont {Novoselov},\ and\
  \citenamefont {Casiraghi}}]{Eckmann2012}%
  \BibitemOpen
  \bibfield  {author} {\bibinfo {author} {\bibfnamefont {A.}~\bibnamefont
  {Eckmann}}, \bibinfo {author} {\bibfnamefont {A.}~\bibnamefont {Felten}},
  \bibinfo {author} {\bibfnamefont {A.}~\bibnamefont {Mishchenko}}, \bibinfo
  {author} {\bibfnamefont {L.}~\bibnamefont {Britnell}}, \bibinfo {author}
  {\bibfnamefont {R.}~\bibnamefont {Krupke}}, \bibinfo {author} {\bibfnamefont
  {K.~S.}\ \bibnamefont {Novoselov}}, \ and\ \bibinfo {author} {\bibfnamefont
  {C.}~\bibnamefont {Casiraghi}},\ }\href@noop {} {\bibfield  {journal}
  {\bibinfo  {journal} {Nano letters}\ }\textbf {\bibinfo {volume} {12}},\
  \bibinfo {pages} {3925} (\bibinfo {year} {2012})}\BibitemShut {NoStop}%
\bibitem [{\citenamefont {Lindahl}\ \emph {et~al.}(2012)\citenamefont
  {Lindahl}, \citenamefont {Midtvedt}, \citenamefont {Svensson}, \citenamefont
  {Nerushev}, \citenamefont {Lindvall}, \citenamefont {Isacsson},\ and\
  \citenamefont {Campbell}}]{Lindahl2012}%
  \BibitemOpen
  \bibfield  {author} {\bibinfo {author} {\bibfnamefont {N.}~\bibnamefont
  {Lindahl}}, \bibinfo {author} {\bibfnamefont {D.}~\bibnamefont {Midtvedt}},
  \bibinfo {author} {\bibfnamefont {J.}~\bibnamefont {Svensson}}, \bibinfo
  {author} {\bibfnamefont {O.~A.}\ \bibnamefont {Nerushev}}, \bibinfo {author}
  {\bibfnamefont {N.}~\bibnamefont {Lindvall}}, \bibinfo {author}
  {\bibfnamefont {A.}~\bibnamefont {Isacsson}}, \ and\ \bibinfo {author}
  {\bibfnamefont {E.~E.}\ \bibnamefont {Campbell}},\ }\href {\doibase
  10.1021/nl301080v} {\bibfield  {journal} {\bibinfo  {journal} {Nano Lett}\
  }\textbf {\bibinfo {volume} {12}},\ \bibinfo {pages} {3526} (\bibinfo {year}
  {2012})}\BibitemShut {NoStop}%
\bibitem [{\citenamefont {Sajadi}\ \emph {et~al.}(2017)\citenamefont {Sajadi},
  \citenamefont {Alijani}, \citenamefont {Davidovikj}, \citenamefont {Goosen},
  \citenamefont {Steeneken},\ and\ \citenamefont {van
  Keulen}}]{SajadiJAP2017}%
  \BibitemOpen
  \bibfield  {author} {\bibinfo {author} {\bibfnamefont {B.}~\bibnamefont
  {Sajadi}}, \bibinfo {author} {\bibfnamefont {F.}~\bibnamefont {Alijani}},
  \bibinfo {author} {\bibfnamefont {D.}~\bibnamefont {Davidovikj}}, \bibinfo
  {author} {\bibfnamefont {J.}~\bibnamefont {Goosen}}, \bibinfo {author}
  {\bibfnamefont {P.~G.}\ \bibnamefont {Steeneken}}, \ and\ \bibinfo {author}
  {\bibfnamefont {F.}~\bibnamefont {van Keulen}},\ }\href@noop {} {\bibfield
  {journal} {\bibinfo  {journal} {Journal of Applied Physics}\ }\textbf
  {\bibinfo {volume} {122}},\ \bibinfo {pages} {234302} (\bibinfo {year}
  {2017})}\BibitemShut {NoStop}%
\bibitem [{\citenamefont {Davidovikj}\ \emph {et~al.}(2017)\citenamefont
  {Davidovikj}, \citenamefont {Alijani}, \citenamefont {Cartamil-Bueno},
  \citenamefont {van~der Zant}, \citenamefont {Amabili},\ and\ \citenamefont
  {Steeneken}}]{Farbod2017}%
  \BibitemOpen
  \bibfield  {author} {\bibinfo {author} {\bibfnamefont {D.}~\bibnamefont
  {Davidovikj}}, \bibinfo {author} {\bibfnamefont {F.}~\bibnamefont {Alijani}},
  \bibinfo {author} {\bibfnamefont {S.~J.}\ \bibnamefont {Cartamil-Bueno}},
  \bibinfo {author} {\bibfnamefont {H.~S.}\ \bibnamefont {van~der Zant}},
  \bibinfo {author} {\bibfnamefont {M.}~\bibnamefont {Amabili}}, \ and\
  \bibinfo {author} {\bibfnamefont {P.~G.}\ \bibnamefont {Steeneken}},\ }\href
  {\doibase 10.1038/s41467-017-01351-4} {\bibfield  {journal} {\bibinfo
  {journal} {nature communications}\ }\textbf {\bibinfo {volume} {8.1}}
  (\bibinfo {year} {2017}),\ 10.1038/s41467-017-01351-4}\BibitemShut {NoStop}%
\bibitem [{\citenamefont {Nicklow}\ \emph {et~al.}(1972)\citenamefont
  {Nicklow}, \citenamefont {Wakabayashi},\ and\ \citenamefont
  {Smith}}]{Nicklow1972}%
  \BibitemOpen
  \bibfield  {author} {\bibinfo {author} {\bibfnamefont {R.}~\bibnamefont
  {Nicklow}}, \bibinfo {author} {\bibfnamefont {N.}~\bibnamefont
  {Wakabayashi}}, \ and\ \bibinfo {author} {\bibfnamefont {H.}~\bibnamefont
  {Smith}},\ }\href@noop {} {\bibfield  {journal} {\bibinfo  {journal}
  {Physical Review B}\ }\textbf {\bibinfo {volume} {5}},\ \bibinfo {pages}
  {4951} (\bibinfo {year} {1972})}\BibitemShut {NoStop}%
\bibitem [{\citenamefont {Blees}\ \emph {et~al.}(2015)\citenamefont {Blees},
  \citenamefont {Barnard}, \citenamefont {Rose}, \citenamefont {Roberts},
  \citenamefont {McGill}, \citenamefont {Huang}, \citenamefont {Ruyack},
  \citenamefont {Kevek}, \citenamefont {Kobrin},\ and\ \citenamefont
  {Muller}}]{Blees2015}%
  \BibitemOpen
  \bibfield  {author} {\bibinfo {author} {\bibfnamefont {M.~K.}\ \bibnamefont
  {Blees}}, \bibinfo {author} {\bibfnamefont {A.~W.}\ \bibnamefont {Barnard}},
  \bibinfo {author} {\bibfnamefont {P.~A.}\ \bibnamefont {Rose}}, \bibinfo
  {author} {\bibfnamefont {S.~P.}\ \bibnamefont {Roberts}}, \bibinfo {author}
  {\bibfnamefont {K.~L.}\ \bibnamefont {McGill}}, \bibinfo {author}
  {\bibfnamefont {P.~Y.}\ \bibnamefont {Huang}}, \bibinfo {author}
  {\bibfnamefont {A.~R.}\ \bibnamefont {Ruyack}}, \bibinfo {author}
  {\bibfnamefont {J.~W.}\ \bibnamefont {Kevek}}, \bibinfo {author}
  {\bibfnamefont {B.}~\bibnamefont {Kobrin}}, \ and\ \bibinfo {author}
  {\bibfnamefont {D.~A.}\ \bibnamefont {Muller}},\ }\href@noop {} {\bibfield
  {journal} {\bibinfo  {journal} {Nature}\ }\textbf {\bibinfo {volume} {524}},\
  \bibinfo {pages} {204} (\bibinfo {year} {2015})}\BibitemShut {NoStop}%
\bibitem [{\citenamefont {Wei}\ \emph {et~al.}(2012)\citenamefont {Wei},
  \citenamefont {Wang}, \citenamefont {Wu}, \citenamefont {Yang},\ and\
  \citenamefont {Dunn}}]{Yujie2012}%
  \BibitemOpen
  \bibfield  {author} {\bibinfo {author} {\bibfnamefont {Y.}~\bibnamefont
  {Wei}}, \bibinfo {author} {\bibfnamefont {B.}~\bibnamefont {Wang}}, \bibinfo
  {author} {\bibfnamefont {J.}~\bibnamefont {Wu}}, \bibinfo {author}
  {\bibfnamefont {R.}~\bibnamefont {Yang}}, \ and\ \bibinfo {author}
  {\bibfnamefont {M.~L.}\ \bibnamefont {Dunn}},\ }\href@noop {} {\bibfield
  {journal} {\bibinfo  {journal} {Nano letters}\ }\textbf {\bibinfo {volume}
  {13}},\ \bibinfo {pages} {26} (\bibinfo {year} {2012})}\BibitemShut {NoStop}%
\bibitem [{\citenamefont {Tersoff}(1992)}]{Tersoff1992}%
  \BibitemOpen
  \bibfield  {author} {\bibinfo {author} {\bibfnamefont {J.}~\bibnamefont
  {Tersoff}},\ }\href {\doibase 10.1103/PhysRevB.46.15546} {\bibfield
  {journal} {\bibinfo  {journal} {Physical Review B}\ }\textbf {\bibinfo
  {volume} {46}},\ \bibinfo {pages} {15546} (\bibinfo {year}
  {1992})}\BibitemShut {NoStop}%
\bibitem [{\citenamefont {Lu}\ \emph {et~al.}(2009)\citenamefont {Lu},
  \citenamefont {Arroyo},\ and\ \citenamefont {Huang}}]{lu2009}%
  \BibitemOpen
  \bibfield  {author} {\bibinfo {author} {\bibfnamefont {Q.}~\bibnamefont
  {Lu}}, \bibinfo {author} {\bibfnamefont {M.}~\bibnamefont {Arroyo}}, \ and\
  \bibinfo {author} {\bibfnamefont {R.}~\bibnamefont {Huang}},\ }\href@noop {}
  {\bibfield  {journal} {\bibinfo  {journal} {Journal of Physics D: Applied
  Physics}\ }\textbf {\bibinfo {volume} {42}},\ \bibinfo {pages} {102002}
  (\bibinfo {year} {2009})}\BibitemShut {NoStop}%
\bibitem [{\citenamefont {Sánchez-Portal}\ \emph {et~al.}(1999)\citenamefont
  {Sánchez-Portal}, \citenamefont {Artacho}, \citenamefont {Soler},
  \citenamefont {Rubio},\ and\ \citenamefont {Ordejón}}]{Sanchez1999}%
  \BibitemOpen
  \bibfield  {author} {\bibinfo {author} {\bibfnamefont {D.}~\bibnamefont
  {Sánchez-Portal}}, \bibinfo {author} {\bibfnamefont {E.}~\bibnamefont
  {Artacho}}, \bibinfo {author} {\bibfnamefont {J.~M.}\ \bibnamefont {Soler}},
  \bibinfo {author} {\bibfnamefont {A.}~\bibnamefont {Rubio}}, \ and\ \bibinfo
  {author} {\bibfnamefont {P.}~\bibnamefont {Ordejón}},\ }\href
  {https://link.aps.org/doi/10.1103/PhysRevB.59.12678} {\bibfield  {journal}
  {\bibinfo  {journal} {Physical Review B}\ }\textbf {\bibinfo {volume} {59}},\
  \bibinfo {pages} {12678} (\bibinfo {year} {1999})}\BibitemShut {NoStop}%
\bibitem [{\citenamefont {Brenner}(1990)}]{Brenner1990}%
  \BibitemOpen
  \bibfield  {author} {\bibinfo {author} {\bibfnamefont {D.~W.}\ \bibnamefont
  {Brenner}},\ }\href@noop {} {\bibfield  {journal} {\bibinfo  {journal}
  {Physical Review B}\ }\textbf {\bibinfo {volume} {42}},\ \bibinfo {pages}
  {9458} (\bibinfo {year} {1990})}\BibitemShut {NoStop}%
\bibitem [{\citenamefont {Brenner}\ \emph {et~al.}(2002)\citenamefont
  {Brenner}, \citenamefont {Shenderova}, \citenamefont {Harrison},
  \citenamefont {Stuart}, \citenamefont {Ni},\ and\ \citenamefont
  {Sinnott}}]{Brenner2002}%
  \BibitemOpen
  \bibfield  {author} {\bibinfo {author} {\bibfnamefont {D.~W.}\ \bibnamefont
  {Brenner}}, \bibinfo {author} {\bibfnamefont {O.~A.}\ \bibnamefont
  {Shenderova}}, \bibinfo {author} {\bibfnamefont {J.~A.}\ \bibnamefont
  {Harrison}}, \bibinfo {author} {\bibfnamefont {S.~J.}\ \bibnamefont
  {Stuart}}, \bibinfo {author} {\bibfnamefont {B.}~\bibnamefont {Ni}}, \ and\
  \bibinfo {author} {\bibfnamefont {S.~B.}\ \bibnamefont {Sinnott}},\
  }\href@noop {} {\bibfield  {journal} {\bibinfo  {journal} {Journal of
  Physics: Condensed Matter}\ }\textbf {\bibinfo {volume} {14}},\ \bibinfo
  {pages} {783} (\bibinfo {year} {2002})}\BibitemShut {NoStop}%
\bibitem [{\citenamefont {Koskinen}\ and\ \citenamefont
  {Kit}(2010)}]{Koskinen2010}%
  \BibitemOpen
  \bibfield  {author} {\bibinfo {author} {\bibfnamefont {P.}~\bibnamefont
  {Koskinen}}\ and\ \bibinfo {author} {\bibfnamefont {O.~O.}\ \bibnamefont
  {Kit}},\ }\href@noop {} {\bibfield  {journal} {\bibinfo  {journal} {Physical
  Review B}\ }\textbf {\bibinfo {volume} {82}},\ \bibinfo {pages} {235420}
  (\bibinfo {year} {2010})}\BibitemShut {NoStop}%
\bibitem [{\citenamefont {Kudin}\ \emph {et~al.}(2001)\citenamefont {Kudin},
  \citenamefont {Scuseria},\ and\ \citenamefont {Yakobson}}]{Kudin2001}%
  \BibitemOpen
  \bibfield  {author} {\bibinfo {author} {\bibfnamefont {K.~N.}\ \bibnamefont
  {Kudin}}, \bibinfo {author} {\bibfnamefont {G.~E.}\ \bibnamefont {Scuseria}},
  \ and\ \bibinfo {author} {\bibfnamefont {B.~I.}\ \bibnamefont {Yakobson}},\
  }\href@noop {} {\bibfield  {journal} {\bibinfo  {journal} {Physical Review
  B}\ }\textbf {\bibinfo {volume} {64}},\ \bibinfo {pages} {235406} (\bibinfo
  {year} {2001})}\BibitemShut {NoStop}%
\bibitem [{\citenamefont {Roldán}\ \emph {et~al.}(2011)\citenamefont
  {Roldán}, \citenamefont {Fasolino}, \citenamefont {Zakharchenko},\ and\
  \citenamefont {Katsnelson}}]{Katsnelson2011}%
  \BibitemOpen
  \bibfield  {author} {\bibinfo {author} {\bibfnamefont {R.}~\bibnamefont
  {Roldán}}, \bibinfo {author} {\bibfnamefont {A.}~\bibnamefont {Fasolino}},
  \bibinfo {author} {\bibfnamefont {K.~V.}\ \bibnamefont {Zakharchenko}}, \
  and\ \bibinfo {author} {\bibfnamefont {M.~I.}\ \bibnamefont {Katsnelson}},\
  }\href@noop {} {\bibfield  {journal} {\bibinfo  {journal} {Physical Review
  B}\ }\textbf {\bibinfo {volume} {83}},\ \bibinfo {pages} {174104} (\bibinfo
  {year} {2011})}\BibitemShut {NoStop}%
\bibitem [{\citenamefont {Nelson}\ and\ \citenamefont
  {Peliti}(1987)}]{Nelson1987}%
  \BibitemOpen
  \bibfield  {author} {\bibinfo {author} {\bibfnamefont {D.}~\bibnamefont
  {Nelson}}\ and\ \bibinfo {author} {\bibfnamefont {L.}~\bibnamefont
  {Peliti}},\ }\href@noop {} {\bibfield  {journal} {\bibinfo  {journal} {J.
  Phys.(Paris)}\ }\textbf {\bibinfo {volume} {48}},\ \bibinfo {pages} {1085}
  (\bibinfo {year} {1987})}\BibitemShut {NoStop}%
\bibitem [{\citenamefont {Nicholl}\ \emph {et~al.}(2015)\citenamefont
  {Nicholl}, \citenamefont {Conley}, \citenamefont {Lavrik}, \citenamefont
  {Vlassiouk}, \citenamefont {Puzyrev}, \citenamefont {Sreenivas},
  \citenamefont {Pantelides},\ and\ \citenamefont {Bolotin}}]{Nicholl2015}%
  \BibitemOpen
  \bibfield  {author} {\bibinfo {author} {\bibfnamefont {R.~J.}\ \bibnamefont
  {Nicholl}}, \bibinfo {author} {\bibfnamefont {H.~J.}\ \bibnamefont {Conley}},
  \bibinfo {author} {\bibfnamefont {N.~V.}\ \bibnamefont {Lavrik}}, \bibinfo
  {author} {\bibfnamefont {I.}~\bibnamefont {Vlassiouk}}, \bibinfo {author}
  {\bibfnamefont {Y.~S.}\ \bibnamefont {Puzyrev}}, \bibinfo {author}
  {\bibfnamefont {V.~P.}\ \bibnamefont {Sreenivas}}, \bibinfo {author}
  {\bibfnamefont {S.~T.}\ \bibnamefont {Pantelides}}, \ and\ \bibinfo {author}
  {\bibfnamefont {K.~I.}\ \bibnamefont {Bolotin}},\ }\href@noop {} {\bibfield
  {journal} {\bibinfo  {journal} {Nature communications}\ }\textbf {\bibinfo
  {volume} {6}} (\bibinfo {year} {2015})}\BibitemShut {NoStop}%
\bibitem [{\citenamefont {Deng}\ and\ \citenamefont {Berry}(2016)}]{Deng2016}%
  \BibitemOpen
  \bibfield  {author} {\bibinfo {author} {\bibfnamefont {S.}~\bibnamefont
  {Deng}}\ and\ \bibinfo {author} {\bibfnamefont {V.}~\bibnamefont {Berry}},\
  }\href@noop {} {\bibfield  {journal} {\bibinfo  {journal} {Materials Today}\
  }\textbf {\bibinfo {volume} {19}},\ \bibinfo {pages} {197} (\bibinfo {year}
  {2016})}\BibitemShut {NoStop}%
\bibitem [{\citenamefont {Gao}\ and\ \citenamefont {Huang}(2014)}]{Gao2014}%
  \BibitemOpen
  \bibfield  {author} {\bibinfo {author} {\bibfnamefont {W.}~\bibnamefont
  {Gao}}\ and\ \bibinfo {author} {\bibfnamefont {R.}~\bibnamefont {Huang}},\
  }\href@noop {} {\bibfield  {journal} {\bibinfo  {journal} {Journal of the
  Mechanics and Physics of Solids}\ }\textbf {\bibinfo {volume} {66}},\
  \bibinfo {pages} {42} (\bibinfo {year} {2014})}\BibitemShut {NoStop}%
\bibitem [{\citenamefont {Bao}\ \emph {et~al.}(2009)\citenamefont {Bao},
  \citenamefont {Miao}, \citenamefont {Chen}, \citenamefont {Zhang},
  \citenamefont {Jang}, \citenamefont {Dames},\ and\ \citenamefont
  {Lau}}]{bao2009}%
  \BibitemOpen
  \bibfield  {author} {\bibinfo {author} {\bibfnamefont {W.}~\bibnamefont
  {Bao}}, \bibinfo {author} {\bibfnamefont {F.}~\bibnamefont {Miao}}, \bibinfo
  {author} {\bibfnamefont {Z.}~\bibnamefont {Chen}}, \bibinfo {author}
  {\bibfnamefont {H.}~\bibnamefont {Zhang}}, \bibinfo {author} {\bibfnamefont
  {W.}~\bibnamefont {Jang}}, \bibinfo {author} {\bibfnamefont {C.}~\bibnamefont
  {Dames}}, \ and\ \bibinfo {author} {\bibfnamefont {C.~N.}\ \bibnamefont
  {Lau}},\ }\href@noop {} {\bibfield  {journal} {\bibinfo  {journal} {Nature
  nanotechnology}\ }\textbf {\bibinfo {volume} {4}},\ \bibinfo {pages} {562}
  (\bibinfo {year} {2009})}\BibitemShut {NoStop}%
\bibitem [{\citenamefont {Wan}\ \emph {et~al.}(2017)\citenamefont {Wan},
  \citenamefont {Nelson},\ and\ \citenamefont {Bowick}}]{Duanduan2017}%
  \BibitemOpen
  \bibfield  {author} {\bibinfo {author} {\bibfnamefont {D.}~\bibnamefont
  {Wan}}, \bibinfo {author} {\bibfnamefont {D.~R.}\ \bibnamefont {Nelson}}, \
  and\ \bibinfo {author} {\bibfnamefont {M.~J.}\ \bibnamefont {Bowick}},\
  }\href@noop {} {\bibfield  {journal} {\bibinfo  {journal} {Physical Review
  B}\ }\textbf {\bibinfo {volume} {96}},\ \bibinfo {pages} {014106} (\bibinfo
  {year} {2017})}\BibitemShut {NoStop}%
\bibitem [{\citenamefont {Plimpton}\ \emph {et~al.}(2007)\citenamefont
  {Plimpton}, \citenamefont {Crozier},\ and\ \citenamefont
  {Thompson}}]{LAMMPS}%
  \BibitemOpen
  \bibfield  {author} {\bibinfo {author} {\bibfnamefont {S.}~\bibnamefont
  {Plimpton}}, \bibinfo {author} {\bibfnamefont {P.}~\bibnamefont {Crozier}}, \
  and\ \bibinfo {author} {\bibfnamefont {A.}~\bibnamefont {Thompson}},\
  }\href@noop {} {\bibfield  {journal} {\bibinfo  {journal} {Sandia National
  Laboratories}\ }\textbf {\bibinfo {volume} {18}},\ \bibinfo {pages} {43}
  (\bibinfo {year} {2007})}\BibitemShut {NoStop}%
\bibitem [{\citenamefont {Tersoff}(1988)}]{Tersoff1988}%
  \BibitemOpen
  \bibfield  {author} {\bibinfo {author} {\bibfnamefont {J.}~\bibnamefont
  {Tersoff}},\ }\href@noop {} {\bibfield  {journal} {\bibinfo  {journal}
  {Physical Review Letters}\ }\textbf {\bibinfo {volume} {61}},\ \bibinfo
  {pages} {2879} (\bibinfo {year} {1988})}\BibitemShut {NoStop}%
\bibitem [{\citenamefont {Klessig}\ and\ \citenamefont
  {Polak}(1972)}]{Polak1972}%
  \BibitemOpen
  \bibfield  {author} {\bibinfo {author} {\bibfnamefont {R.}~\bibnamefont
  {Klessig}}\ and\ \bibinfo {author} {\bibfnamefont {E.}~\bibnamefont
  {Polak}},\ }\href@noop {} {\bibfield  {journal} {\bibinfo  {journal} {SIAM
  Journal on Control}\ }\textbf {\bibinfo {volume} {10}},\ \bibinfo {pages}
  {524} (\bibinfo {year} {1972})}\BibitemShut {NoStop}%
\bibitem [{\citenamefont {Evans}\ and\ \citenamefont
  {Holian}(1985)}]{Evans1985}%
  \BibitemOpen
  \bibfield  {author} {\bibinfo {author} {\bibfnamefont {D.~J.}\ \bibnamefont
  {Evans}}\ and\ \bibinfo {author} {\bibfnamefont {B.~L.}\ \bibnamefont
  {Holian}},\ }\href@noop {} {\bibfield  {journal} {\bibinfo  {journal} {The
  Journal of chemical physics}\ }\textbf {\bibinfo {volume} {83}},\ \bibinfo
  {pages} {4069} (\bibinfo {year} {1985})}\BibitemShut {NoStop}%
\bibitem [{\citenamefont {Rao}(2007)}]{Rao2007}%
  \BibitemOpen
  \bibfield  {author} {\bibinfo {author} {\bibfnamefont {S.~S.}\ \bibnamefont
  {Rao}},\ }\href@noop {} {\emph {\bibinfo {title} {Vibration of continuous
  systems}}}\ (\bibinfo  {publisher} {John Wiley \& Sons},\ \bibinfo {year}
  {2007})\BibitemShut {NoStop}%
\bibitem [{SM()}]{SM}%
  \BibitemOpen
  \href@noop {} {\emph {\bibinfo {title} {See {S}upplemental {M}aterial at
  http://link.aps.org/supplemental/XXXX for more details about obtaining the
  equations of motion in continuum framework}}}\BibitemShut {NoStop}%
\bibitem [{\citenamefont {Amabili}(2008)}]{Amabili2008}%
  \BibitemOpen
  \bibfield  {author} {\bibinfo {author} {\bibfnamefont {M.}~\bibnamefont
  {Amabili}},\ }\href {\doibase 10.1017/CBO9780511619694} {\emph {\bibinfo
  {title} {Nonlinear vibrations and stability of shells and plates}}}\
  (\bibinfo  {publisher} {Cambridge University Press},\ \bibinfo {year}
  {2008})\BibitemShut {NoStop}%
\bibitem [{\citenamefont {Mounet}\ and\ \citenamefont
  {Marzari}(2005)}]{mounet2005first}%
  \BibitemOpen
  \bibfield  {author} {\bibinfo {author} {\bibfnamefont {N.}~\bibnamefont
  {Mounet}}\ and\ \bibinfo {author} {\bibfnamefont {N.}~\bibnamefont
  {Marzari}},\ }\href@noop {} {\bibfield  {journal} {\bibinfo  {journal}
  {Physical Review B}\ }\textbf {\bibinfo {volume} {71}},\ \bibinfo {pages}
  {205214} (\bibinfo {year} {2005})}\BibitemShut {NoStop}%
\end{thebibliography}

%
\end{document}